\documentclass[aps,prd,11pt,nofootinbib,tightenlines,preprintnumbers,superscriptaddress,notitlepage]{revtex4-2}
\usepackage[a4paper,left=2.5cm,right=2.5cm, top=2.0cm,bottom=2.0cm]{geometry}
\usepackage[english]{babel}
\usepackage[applemac]{inputenc}
\usepackage[colorlinks,citecolor=blue,linktoc=all,linkcolor=black,urlcolor=black]{hyperref}
\usepackage{amsfonts,amsmath,amssymb}
\usepackage{graphicx}
\usepackage{booktabs, multirow}
\usepackage[detect-all]{siunitx}
\usepackage{hyperxmp}
\usepackage{color}

\newcommand{\Fig}[1]{Fig.~\ref{#1}}
\newcommand{\Eq}[1]{Eq.~(\ref{#1})}
\newcommand{\Section}[1]{Section~\ref{#1}}
\newcommand{\Table}[1]{Table~\ref{#1}}

\newcommand{\dof}{\mathrm{d.o.f.}}
\newcommand{\fm}{\mathrm{fm}}
\newcommand{\MeV}{\mathrm{MeV}}

\newcommand{\psib}{\overline{\psi}}	
\newcommand{\xv}{\vec{x}}

\newcommand{\pv}{\vec{p}}
\newcommand{\chib}{\overline{\chi}}

\newcommand{\Tr}{\mathrm{Tr}}
\newcommand{\Nev}{N_{\mathrm{ev}}}
\newcommand{\etap}{\eta^{\prime}}
\newcommand{\ubar}{\bar{u}}
\newcommand{\dbar}{\bar{d}}
\newcommand{\sbar}{\bar{s}}
\newcommand{\ndof}{\mathrm{d.o.f.}}

\makeatother

\begin{document}
%
\title{Lattice QCD calculation of the $\eta$ and $\etap$ meson masses at the physical point using rooted staggered fermions}
\author{Willem E. A. Verplanke}
\affiliation{Aix-Marseille Universit\'e, Universit\'e de Toulon, CNRS, CPT, Marseille, France}
\author{Zoltan Fodor} 
\affiliation{Department of Physics, University of Wuppertal, D-42119 Wuppertal, Germany} 
\affiliation{J\"ulich Supercomputing Centre, Forschungszentrum J\"ulich, D-52428 J\"ulich, Germany} 
\affiliation{Institute for Theoretical Physics, E\"otv\"os University, H-1117 Budapest, Hungary} 
\affiliation{Physics Department, Pennsylvania State University, University Park, PA 16802, USA} 
\author{Antoine G\'erardin}
\email{antoine.gerardin@cpt.univ-mrs.fr}
\affiliation{Aix-Marseille Universit\'e, Universit\'e de Toulon, CNRS, CPT, Marseille, France}
\author{Jana N. Guenther} 
\affiliation{Department of Physics, University of Wuppertal, D-42119 Wuppertal, Germany}
\author{Laurent Lellouch} 
\affiliation{Aix-Marseille Universit\'e, Universit\'e de Toulon, CNRS, CPT, Marseille, France}
\author{Kalman K. Szabo}
\affiliation{J\"ulich Supercomputing Centre, Forschungszentrum J\"ulich, D-52428 J\"ulich, Germany}
\author{Balint C. Toth}
\affiliation{Department of Physics, University of Wuppertal, D-42119 Wuppertal, Germany}
\affiliation{J\"ulich Supercomputing Centre, Forschungszentrum J\"ulich, D-52428 J\"ulich, Germany}
\author{Lukas Varnhorst}
\affiliation{Department of Physics, University of Wuppertal, D-42119 Wuppertal, Germany}

\begin{abstract}
We present a lattice calculation of the $\eta$ and $\eta^{\prime}$ meson masses at the physical point and in the continuum limit, based on $N_f = 2+1+1$ flavors of rooted staggered quarks. 
Our analysis includes gauge ensembles at the physical pion and kaon masses spread over six lattice spacings in the range [0.064-0.1315]~fm. 
Our main results read $m_{\eta} = 543.5(5.6)~$MeV and $m_{\etap} = 986(38)~$MeV, consistent with the experimental values. 
This is an important numerical test that supports the validity of the fourth root procedure used in the staggered quark formalism. 
This calculation was the first step towards extracting the pseudoscalar transition form factors of the $\eta$ and $\etap$ mesons that play a crucial role in the hadronic light-by-light contribution to the muon $g-2$. 
\end{abstract}
\maketitle

\section{Introduction \label{sec:introduction} }

In the chiral limit, where up, down and strange quark masses are set to zero, the classical Lagrangian of QCD exhibits an $SU(3)_L \times SU(3)_R \times U(1)_A \times U(1)_B$ global symmetry. 
In Nature, the $SU(3)_L \times SU(3)_R$ symmetry is spontaneously broken down to $SU(3)_V$, leading to eight Goldstone bosons, one for each broken generator. 
These Goldstone bosons can be identified with the octet of light pseudoscalar mesons $(\pi^0, \pi^{\pm}, K^0, \bar{K}^0, K^{\pm},\eta$) and their different masses can be explained by the non-vanishing quark-mass matrix that explicitly breaks $SU(3)_f$ flavor symmetry. 
On the other hand, the $U(1)_A$ symmetry is anomalous~\cite{Adler:1969gk,Bell:1969ts} and the flavor-singlet $\etap$ acquires a mass, even in the chiral limit. The origin of the singlet meson mass can be explained by the non-trivial topological structure of QCD and the existence of instantons~\cite{Belavin:1975fg,tHooft:1976rip}.  

Flavor symmetry breaking also leads to mixing among the states with the same quantum numbers. The three neutral mesons $\pi_3$, $\eta_0$ and $\eta_8$, defined in the chiral limit and with quantum numbers $J^{PC} = 0^{-+}$, mix to give the physical $\pi^0$, $\eta$ and $\etap$ mesons observed experimentally. 
The mixing between the pion and the other two neutral mesons, proportional to $m_u-m_d$, is small and vanishes in the isospin limit. 
For the $\eta$ and $\etap$, $SU(3)_f$ breaking effects are more important and both mesons can be seen as a mixture of the $\eta_0$ and $\eta_8$ states. 
Only in the $SU(3)_f$ symmetric limit can one perform the assignment $\eta=\eta_8$ and $\etap=\eta_0$. 
In addition, one would expect mixing with glueballs with quantum numbers $J^{PC} = 0^{-+}$. 
However, due to their large masses, above 2.5~GeV as suggested by lattice calculations~\cite{Chen:2021dvn,Vadacchino:2023vnc}, this mixing is not considered in this study. 

Reproducing the mass of the $\eta$ and $\etap$ mesons from first-principle using lattice QCD simulations is an important test of our understanding of the chiral symmetry breaking mechanism and the complex topological structure of QCD~\cite{Witten:1979vv}. 
However lattice calculations of the flavor-singlet meson masses are challenging due to the presence of large quark-disconnected contributions and large autocorrelation times, especially for the $\etap$ meson. 
The spectrum has been studied for a long time on the lattice using $N_f = 2+1$ domain-wall fermions~\cite{Christ:2010dd,Fukaya:2015ara}, $N_f = 2+1+1$ twisted-mass fermions~\cite{Ottnad:2012fv,Ottnad:2015hva,Ottnad:2017bjt} and $N_f = 2+1$ Wilson-clover quarks~\cite{Dudek:2013yja,Bali:2014pva,Bali:2021qem}.  
However, the spectrum has never been computed directly at the physical pion mass and the results quoted at the physical point rely on a chiral extrapolation. 
There is also less literature for rooted staggered quarks despite its relevance in assessing the validity of the rooting procedure and the ability of staggered quarks to reproduce the $U(1)_A$ anomaly correctly.
A first attempt was presented in~\cite{Gregory:2011sg} using two gauge ensembles with different pion masses and lattice spacings. The result suggests a good agreement with experiment but  lacks a proper extrapolation to the physical point. In~\cite{Durr:2012te}, the validity of rooting has been investigated through the massive Schwinger model with a single flavor where the corresponding $\eta$ meson acquires a mass through the axial anomaly. 
Finally, in~\cite{Donald:2011if}, the authors analyzed the behavior of flavor-singlet and taste-singlet correlators by looking at the near-zero modes of the Dirac operator.

Staggered transformations~\cite{Kogut:1974ag} can be used to diagonalize, in spinor space, the naive discretization of the Dirac operator in any background gauge field (see~\cite{Follana:2006rc,Golterman:2024xos} for a review). For each flavor, the resulting single-component spinor field reduces the number of doublers from 16 to 4, called tastes, and allows for much faster numerical implementations at the cost of having extra tastes that need to be eliminated.  
In the continuum limit, all tastes are expected to be equivalent and we are left with an exact fourfold degeneracy of quarks per staggered fermion field. 
In practice, to further reduce the number of tastes from four to one in the sea, one usually takes the fourth root of the fermion determinant, for each flavor introduced in the simulation. 
This method, called \textit{rooting}, is justified in the continuum limit due to the exact fourfold degeneracy~\cite{Sharpe:2006re} such that the determinant of each flavor factorizes into four identical determinants representing a single fermion. 
However, this step is more controversial at finite lattice spacings where taste violation effects break the degeneracy among tastes, leading to a non-unitary theory~\cite{Creutz:2007rk,Creutz:2007yg}. These unphysical effects are, however, expected to vanish in the continuum limit~\cite{Sharpe:2006re,Bernard:2006zw,Kronfeld:2007ek,Golterman:2008gt,Adams:2008db}. Since most arguments against rooting involve the axial anomaly, the $\etap$ meson plays a central role. 

This work is also a first step toward the lattice calculation of the pseudoscalar transition form factors~\cite{Gerardin:2023naa} that are key inputs to evaluate the hadronic light-by-light scattering contribution in the muon $g-2$~\cite{Gerardin:2020gpp,Aoyama:2020ynm}. 

The rest of this paper is organized as follows. In Section~\ref{sec:methodology}, we present the methodology to extract the $\eta$ and $\etap$ meson masses from lattice QCD simulations. 
In Section~\ref{sec:setup} we discuss the implementation of two staggered taste-singlet pseudoscalar interpolating operators and we present the different noise-reduction techniques that have been implemented to evaluate the noisy quark-disconnected contributions. 
In Section~\ref{sec:pion}, we start with a comparison of the two taste-singlet operators, focusing on the quark-connected contribution. 
In Section~\ref{sec:res}, we present our results for the spectrum on individual gauge ensembles, where different strategies to extract the masses are discussed. Finally, we extrapolate the data to the continuum limit before concluding in Section~\ref{sec:ccl}.

\section{Methodology \label{sec:methodology} }

We work in the isospin limit of QCD defined by $m_l  \equiv m_u = m_d$ and electromagnetic interactions are neglected such that the pion does not mix with the $\eta$ and $\etap$ pseudoscalars. 
In this limit, the $\eta$ meson is a stable particle as the hadronic decays $\eta \to \pi^+ \pi^- \pi^0$ and $\eta \to 3\pi^0$ break isospin symmetry.
However, because the $\etap$ is significantly heavier than the $\eta$, the strong decays $\etap \to \eta \pi^+ \pi^-$ and $\etap \to \eta \pi^0 \pi^0$ are allowed. 
In this paper, we nevertheless treat the $\etap$ meson as a stable particle. This approximation is well justified since the decay width $0.188(6)~$MeV~\cite{ParticleDataGroup:2024cfk} is small compared to the resolution we are aiming for in this work.
The charm quark contribution was shown to be small by the ETM collaboration~\cite{Ottnad:2012fv} and is thus neglected here. 

There is a lot of freedom regarding the choice of interpolating operators, with the only requirement that they should have non-zero overlap with the physical states of interest. For the pion, we consider the flavor structure
\begin{equation}
P_{3}(x) = \frac{1}{\sqrt{2}}\left(\ubar \gamma_5 u(x) - \dbar \gamma_5 d(x) \right) = \psib(x) \frac{\lambda_3}{2} \psi(x) \,,
\end{equation}
where $\psib = (\ubar,\dbar,\sbar)$ is a vector in flavor space and $\lambda_a$ are the Gell-Mann matrices. For the $\eta^{(\prime)}$, a standard choice of interpolating operators is given by the flavor-octet $\eta_8$ and flavor-singlet $\eta_0$ operators 
\begin{subequations}
\begin{align}
P_{8}(x) &= \frac{1}{\sqrt{6}}\left(\ubar \gamma_5 u(x) + \dbar \gamma_5 d(x) - 2 \sbar \gamma_5 s(x)\right) = \psib(x) \frac{\lambda_8}{2} \psi(x),\\
P_{0}(x) &= \frac{1}{\sqrt{3}} \left(\ubar \gamma_5 u(x) + \dbar \gamma_5 d(x) + \sbar \gamma_5 s(x)\right)= \psib(x) \frac{\lambda_0}{2} \psi(x) \,,
\end{align}
\end{subequations}
where $\lambda_0 = \frac{2}{\sqrt{3}} \mathbf{I}$. 
In the SU(3) flavor limit, $\langle 0 | P_8 | \eta_0 \rangle = \langle 0 | P_0 | \eta_8 \rangle = 0$, and the time dependence of the two-point correlation functions built from these operators is governed by the $\eta_8$ and $\eta_0$ mesons at large Euclidean times. 
The lattice implementation of those operators, projected on given spatial momenta, are denoted by $O_3$, $O_8$ and $O_0$ and their specific forms will be discussed in \Section{sec:op}.  
Away from the SU(3) flavor limit, the flavor-singlet and flavor-octet states mix to give the physical $\eta$ and $\etap$ mesons, and we are left with the following correlation matrix
\begin{equation}
\label{eq:C2iso}
C_I(t) = 
\begin{pmatrix}
\langle O_8(t) O^{\dag}_8(0) \rangle & \langle O_8(x) O^{\dag}_0(0) \rangle \\[2mm]
\langle O_0(t) O^{\dag}_8(0) \rangle & \langle O_0(x) O^{\dag}_0(0) \rangle
\end{pmatrix} \,.
\end{equation}
In this isospin basis, the correlators are given by
\begin{subequations}
\begin{align}
\langle O_8(t) O^{\dag}_8(0) \rangle 	&= \frac{1}{3} \left( \mathcal{C}_{l} + 2 \mathcal{C}_{s} + 2 \mathcal{D}_{ll} + 2 \mathcal{D}_{ss} - 2 \mathcal{D}_{ls} - 2 \mathcal{D}_{sl} \right) \\
\langle O_8(t) O^{\dag}_0(0) \rangle &=  \frac{\sqrt{2}}{3} \left( \mathcal{C}_{l} - \mathcal{C}_{s} + 2 \mathcal{D}_{ll} - \mathcal{D}_{ss} + \mathcal{D}_{ls} - 2 \mathcal{D}_{sl} \right)  \\
\langle O_0(t) O^{\dag}_8(0) \rangle 	&=  \frac{\sqrt{2}}{3} \left( \mathcal{C}_{l} - \mathcal{C}_{s} + 2 \mathcal{D}_{ll} - \mathcal{D}_{ss} + \mathcal{D}_{sl} - 2 \mathcal{D}_{ls} \right)  \\
\langle O_0(x) O^{\dag}_0(0) \rangle &= \frac{1}{3} \left( 2\mathcal{C}_{l} + \mathcal{C}_{s} + 4 \mathcal{D}_{ll} + \mathcal{D}_{ss} + 2 \mathcal{D}_{ls} + 2 \mathcal{D}_{sl} \right)
\end{align}
\end{subequations}
where $\mathcal{C}_{f}$ is the quark-connected contribution of flavor $f$ and $\mathcal{D}_{f_1f_2}$ is the disconnected contribution with flavors $f_1$ and $f_2$ in each quark loop (see \Fig{fig:2pt}). 
The observation that this matrix is off-diagonal away from the SU(3)$_f$ symmetric point underlines the fact that the SU(3) octet and singlet states are not eigenstates but mix to give the physical $\eta$ and $\eta^{\prime}$ mesons. Neglecting for a moment higher excited states, that are exponentially suppressed at large Euclidean time, and oscillating contributions from the opposite parity channel that are specific to staggered quarks, the masses can be extracted from the time dependence of the correlation matrix
\begin{equation}
\langle O_a(t) O^{\dag}_b(0) \rangle = \frac{Z^{(a)}_\eta Z^{(b)*}_{\eta}}{2E_{\eta}} e^{-E_{\eta} t} + \frac{Z^{(a)}_{\etap} Z^{(b)*}_{\etap}}{2E_{\etap}} e^{-E_{\etap} t} + \cdots
\label{eq:spectraldec}
\end{equation}
where the four overlap factors are defined by $Z_n^{(i)} = \langle n | O_i | 0 \rangle$. 
On the lattice, it is often more natural to work in the flavor basis
\begin{equation}
P_l(x) = \frac{1}{\sqrt{2}} (  \ubar \gamma_5 u(x) +  \dbar \gamma_5 d(x)) \,, \quad P_s(x) = \sbar \gamma_5 s(x) \,,
\end{equation}
where the correlation matrix, in terms of the corresponding lattice interpolating operators $O_l$ and $O_s$, takes the simpler form
\begin{equation}
C_F(t) = \begin{pmatrix}
\langle O_l(t) O^{\dag}_l(0) \rangle & \langle O_l(x) O^{\dag}_s(0) \rangle \\[2mm]
\langle O_s(t) O^{\dag}_l(0) \rangle & \langle O_s(x) O^{\dag}_s(0) \rangle
\end{pmatrix} =
\begin{pmatrix}
\mathcal{C}_{ll} + 2 \mathcal{D}_{ll} & \sqrt{2} \mathcal{D}_{ls}  \\[2mm]
\sqrt{2} \mathcal{D}_{sl} & \mathcal{C}_{ss} + \mathcal{D}_{ss}
\end{pmatrix} \,.
\label{eq:fbasis}
\end{equation}
Since we are eventually interested in the extraction of the pseudoscalar transition form factors, the mixing between the two mesons is defined in terms of amplitudes for the pseudoscalar densities. In the flavor basis, this mixing can be described in terms of two constants and two angles
\begin{equation}
\begin{pmatrix}
Z_{\eta}^{(l)}  & Z_{\eta}^{(s)} \\
Z_{\etap}^{(l)} & Z_{\etap}^{(s)}
\end{pmatrix} =
\begin{pmatrix}
c_l \cos \phi_l  & -c_s \sin \phi_s \\
c_l \sin \phi_l   & c_s \cos \phi_s 
\end{pmatrix} \,,
\end{equation}
with $\tan \phi_l = Z_{\etap}^{(l)} / Z_{\eta}^{(l)}$, $\tan \phi_s = - Z_{\eta}^{(s)} / Z_{\etap}^{(s)}$.
Due to correlation among data, one expects to have a smaller statistical error on the geometric mean, $\tan^2 \phi_F = - (Z_l^{(\etap)} Z_s^{(\eta)} )( Z_l^{(\eta)} Z_s^{(\etap)} )$, rather than on $\phi_l$ and $\phi_s$ separately~\cite{Michael:2013gka,Ottnad:2015hva}. . 

Finally, in finite volume, the correlation functions can acquire a $t$-independent fully-disconnected contribution due to a partial sampling of the topological charge~\cite{Aoki:2007ka,Bali:2014pva}. Thus, in addition to the correlation matrix itself, we also consider the subtracted correlator $\overline{C}_{ij}(t)  = C_{ij}(t+\Delta t)- C_{ij}(t)$ with $\Delta t/a = 1,2$~\cite{Ottnad:2017bjt,Bali:2021qem}. This subtraction also tends to reduce correlations between consecutive time slices, leading to more stable fits in~\Section{sec:res}.

\section{Lattice setup } \label{sec:setup}

\subsection{Lattice gauge ensembles}

\begin{figure}[t]
	\centering
	\includegraphics[width=0.48\textwidth]{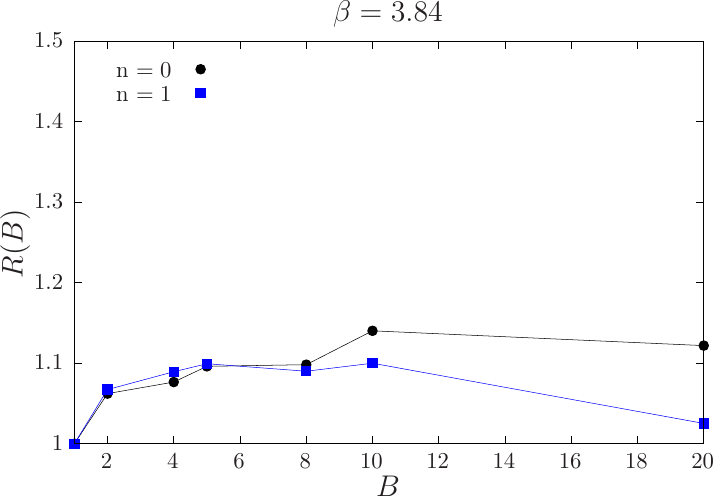}
	\includegraphics[width=0.48\textwidth]{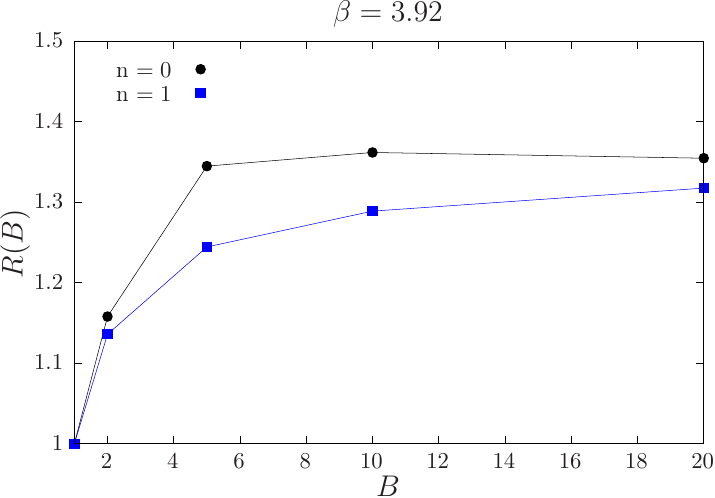}
	\caption{Normalized jackknife error as a function of bin size for the quantity $R(B) = \sum_{t=t_1}^{t_2} \mathcal{D}_{ll}(t/a)$ and for momenta $|\vec{p}| = n(2\pi/L)$, $n=0,1$. We use $t_1 = 2a$ and $t_2 = 8a$. The bin sizes used in the final analysis for these ensembles (in units of configurations separated by 10 trajectories) are 20 and 10 respectively.}
	\label{fig:cor}
\end{figure}

Our calculation is based on a subset of ensembles generated by the Budapest-Marseille-Wuppertal collaboration~\cite{Borsanyi:2020mff}. They have been generated using $N_f = 2+1+1$ dynamical staggered fermions with four steps of stout smearing. The bare quark masses have been tuned such that the Goldstone mesons are at nearly physical pion and kaon mass. The lattice spacing is set using the $\Omega$ baryon mass and we exploit six values of the lattice spacing in the range $[0.0640-0.1315]$~fm to extrapolate our result to the continuum limit. We consider large volumes with $L>6$~fm where finite-volume effects will be shown to be negligible. Simulations are performed in the isospin limit where $m_u = m_d \equiv m_{l}$. More details about these ensembles can be found in~\cite{Borsanyi:2020mff,Budapest-Marseille-Wuppertal:2017okr} and the main properties relevant for this work are summarized in \Table{tab:ens}.

Statistical errors are estimated using the jackknife method after blocking the data over consecutive gauge field configurations. In \Fig{fig:cor} we present the statistical error of the light-quark disconnected correlator $\mathcal{D}_{ll}$ as a function of the bin size. The block size is large enough to essentially suppress all visible autocorrelations. At our finest lattice spacing, the relatively small number of configurations might however hide longer range autocorrelations, especially for the $\etap$ meson. 

\begin{table}[t]
\renewcommand{\arraystretch}{1.1}
\caption{Parameters of the simulations: the bare coupling $\beta = 6/g_0^2$, the lattice resolution, the lattice spacing $a$ and the spatial extent $L$ in physical units, the bare light and strange quark masses and the number of gauge configurations. Only the large-volume ensembles with $L>6~$fm are included in the final analysis. Ensembles with smaller physical extents are given for completeness and were used in~\cite{Gerardin:2023naa}.}
\vskip 0.1in
\begin{tabular}{l@{\hskip 01em}c@{\hskip 01em}l@{\hskip 01em}c@{\hskip 02em}l@{\hskip 01em}l@{\hskip 01em}l@{\hskip 01em}}
\hline
$\quad\beta\quad$	&	$(L/a)^3\times (T/a)$ 	&	$a~[\fm]$	&	$L~[\fm]$	&	$am_l$		&	$am_s$	&	$\#$confs	 \\
\hline
$3.7000$	&	$48^3\times64$	& 	0.1315	& 	6.3	& 0.00205349 & 0.0572911	&  900	\\
		&	$32^3\times64$	& 			& 	4.2	& 0.00205349 & 0.0572911	&  900	\\
		&	$24^3\times48$	& 			&	3.2	& 0.00205349 & 0.0572911	&  700	\\
\hline
$3.7500$	&	$56^3\times96$	& 	0.1191	&	6.7	& 0.00184096 & 0.0495930	&  500	\\
		&	$56^3\times96$	& 			&	6.7	& 0.00176877 & 0.0516173	&  500	\\
		&	$56^3\times96$	& 			&	6.7	& 0.00184096 & 0.0516173	&  500  	\\
\hline
$3.7553$	&	$56^3\times84$	& 	0.1116	&	6.2	& 0.00171008 & 0.0476146	&  500	\\
		&	$56^3\times84$	& 			&	6.2	& 0.00171008 & 0.0485669	&  500	\\
		&	$56^3\times84$	& 			&	6.2	& 0.00174428 & 0.0461862	&  500	\\
		&	$28^3\times56$	& 			&	3.1	& 0.00171008 & 0.0476146	&  850	\\
\hline
$3.8400$	&	$64^3\times96$	& 	0.0952	&	6.1	& 0.00151556 & 0.0431935	&  500 	\\
		&	$64^3\times96$	& 			&	6.1	& 0.001455     & 0.04075		&  1100	\\
		&	$32^3\times64$	& 			&	3.0	& 0.00151556 & 0.0431935	&  1100	\\
		&	$32^3\times64$	& 			&	3.0	& 0.00143       & 0.0431935	&  1050	\\
		&	$32^3\times64$	& 			&	3.0	& 0.001455     & 0.04075		&  1100	\\
		&	$32^3\times64$	& 			&	3.0	& 0.001455     & 0.03913		&  1100	\\
\hline
$3.9200$	&	$80^3\times128$	& 	0.0787	&	6.3	& 0.001172     & 0.03244		&  500 	\\
		&	$80^3\times128$	& 			&	6.3	& 0.0012         & 0.0332856	&  500	\\
		&	$40^3\times80$	& 			&	3.1	& 0.001207     & 0.032		&  550	\\
		&	$40^3\times80$	& 			&	3.1	& 0.0012         & 0.0332856	&  350	\\
\hline
$4.0126$	&	$96^3\times144$	& 	0.0640	&	6.1	& 0.000977     & 0.0264999	&  500 	\\
		&	$96^3\times144$	& 			&	6.1	& 0.001002     & 0.027318		&  450 	\\
		&	$48^3\times96$	& 			&	3.1	& 0.00095897 & 0.0264999	&  850	\\
		&	$48^3\times96$	& 			&	3.1	& 0.001002     & 0.027318		&  450  	\\
\hline
 \end{tabular} 
\label{tab:ens}
\end{table}

\subsection{Conventions}

In this section we summarize the main definitions to set the notations used in this paper. For a review on staggered quarks, we refer the reader to~\cite{Follana:2006rc,Sharpe:2006re}. The staggered Dirac operator for a quark of flavor $f$, that acts on a single component spinor field $\chi$ with a background gauge field $U_{\mu}(x)$ reads
\begin{equation}
D_f(x,y) =  \frac{1}{2a} \sum_{\mu}  \eta_{\mu}(x) \left[ U_{\mu}(x) \delta_{x+a\hat{\mu},y} - U_{\mu}^{\dag}(x-a\hat{\mu}) \delta_{x-a\hat{\mu},y}  \right]  + m_f \delta_{x,y} \,,
\label{eq:D}
\end{equation}
where $a$ is the lattice spacing, $m_f$ the bare quark mass and $\eta_{\mu}(x)$ are phase factors satisfying $\Omega^{\dag}(x) \gamma_{\mu} \Omega(x) = \eta_{\mu}(x)$ with $\Omega(x) =  \gamma_1^{x_1/a}  \gamma_2^{x_2/a}  \gamma_3^{x_3/a}  \gamma_4^{x_4/a}$ and $\gamma_{\mu}$ are the usual gamma matrices. In \Eq{eq:D}, the index $\mu$ runs from 1 to 4 and $\hat{\mu}$ denotes the unit vector in the direction $\mu$. The fermionic action for the flavor $f$ can be written in term of the Dirac operator as 
\begin{equation}
{\mathcal S}_f = a^4 \sum_{x,y} \chib(x) D_f \, \chi(y) \,.
\end{equation}
In the following, we use the conventions where
\begin{equation}
\eta_1(x) = 1 \,,\quad \eta_2(x) = (-1)^{x_1/a} \,,\quad \eta_3(x) = (-1)^{(x_1+x_2)/a} \,,\quad \eta_4(x) = (-1)^{(x_1+x_2+x_3)/a}  \,.
\end{equation}
It is useful to note that $D^{\dag}(y,x) = \epsilon(x) D(x,y) \epsilon(y)$ where $\epsilon = (-1)^{(x_1+x_2+x_3+x_4)/a}$ is the parity factor. This property is similar the $\gamma_5$-hermiticity of the Wilson Dirac propagator and it holds on a single gauge configuration. 

\subsection{Pseudoscalar interpolating operators} \label{sec:op}

\begin{figure}[t]
	\centering
	\includegraphics[width=0.85\textwidth]{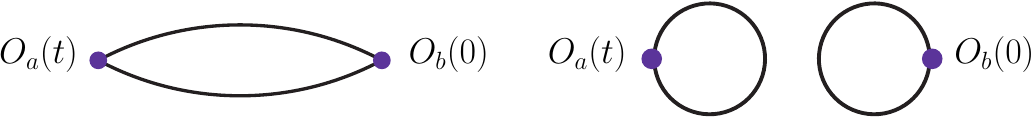}
	\caption{Connected and quark-disconnected contributions to the pseudoscalar two-point correlation functions in~\Eq{eq:C2iso}.}
	\label{fig:2pt}
\end{figure}

The pseudoscalar two-point correlation function involves Wick contractions with quark-disconnected contributions where quark and anti-quarks annihilates into gluons, see \Fig{fig:2pt}. As low-energy gluons are taste singlets, such contributions require the use of taste-singlet pseudoscalar operators~\cite{Donald:2011if}. 
Using the notations of~\cite{Follana:2006rc}, where operators are given in the spin-taste basis as $\Gamma_S \otimes \Gamma_T$, two appropriate choices are given by the 3-link $\gamma_4 \gamma_5 \otimes 1$ and the 4-link $\gamma_5 \otimes 1$ operators~\cite{Altmeyer:1992dd}. 
The 3-link interpolating operator $\hat{O}_3$ is explicitly given by
\begin{equation}
O_3(x) = \frac{1}{6} \sum_{i,j,k} \epsilon_{ijk} \, \chib(x) [ \eta_i \Delta_i [ \eta_j \Delta_j [ \eta_k \Delta_k ] ] ]  \chi(x)  \equiv \chib(x) \hat{O}_3 \chi(x) 
\label{eq:o3}
\end{equation}
where $\Delta_{\mu} \chi(x) = \frac{1}{2} [ U_{\mu}(x) \chi(x+a\hat{\mu}) + U^{\dag}_{\mu}(x-a\hat{\mu}) \chi(x-a\hat{\mu})  ]$ is the symmetric shift operator and $\epsilon_{ijk}$ is the rank-3 Levi-Civita tensor. In order to maintain Lorentz covariance, we average over all equivalent shortest paths in Eq.~(\ref{eq:o3}) with a factor given by the sign of the permutation $(i,j,k)$. Gauge invariance is ensured by the insertion of the gauge field in the symmetric shift. The 4-link operator $\hat{O}_4$ is non-local in time and reads
\begin{equation}
O_4(x) = \frac{1}{2} \eta_4(x) \left[ \chib(x)  \hat{O}_{3} \chi_{+}(x) + \chib_+(x)  \hat{O}_{3} \chi(x) \right] 
\end{equation}
where $\chi_+(x) = U_{4}(x) \chi(x+a\hat{t})$ and $\chib_+(x) = \chib(x+a\hat{t}) U^{\dag}_{4}(x)$. The two-point correlation functions are built using either the 3-link or the 4-link pseudoscalar operator both at the source and at the sink. We have not considered the case where different operators are used at the source and at the sink. A comparison of the results given by the two operators is provided in Section~\ref{sec:opcmp}.

\subsection{Staggered correlation functions}

The light and strange quark-connected contributions are evaluated using $U(1)$ stochastic sources with support on a single time slice (time dilution) and the correlators are explicitly symmetrized in time with respect to $T/2$. 
For the quark-disconnected contributions, several gauge noise reduction techniques have been investigated. 
We have implemented low-mode averaging (LMA)~\cite{DeGrand:2004wh,Giusti:2004yp} where the quark propagator is written as $S_f = S_f^{\rm (lm)} + S_f^{\rm (hm)}$ and
\begin{equation}
S_f^{\rm (lm)}(x,y) = \sum_{n=1}^{\Nev} \frac{ \phi_n(x) \phi_n^{\dag}(y)  }{ \lambda_n + m_f}
\end{equation}
is the low-mode contribution, expressed in term of the $\Nev$ lowest eigenvectors $\phi_k$ and corresponding eigenvalues $\lambda_n$ of the massless Dirac operator. The normalization $\sum_x \phi_n^{\dag}(x) \phi_m(x) = \delta_{n,m}$ is used. This expression provides an exact estimator of the all-to-all propagator in the space spanned by the $N_{\rm ev}$ eigenvectors and volume averaging significantly reduces the statistical uncertainty. 
In practice, we use 1000 eigenvectors of the even-odd Dirac operator on our $L=6$~fm ensembles. For smaller volumes, the number of eigenmodes is roughly rescaled by the volume ratio.  
The high-mode contribution to the quark propagator can be estimated stochastically using a set of $N_s$ random sources $\eta_s$ 
\begin{equation}
S_f^{\rm (hm)}(x,y) = \frac{1}{N_s}  \sum_{s=1}^{N_s} (D^{-1}  P \eta_s)(x) \, \eta^{\dag}_s(y)  \,, \quad P = 1 - \sum_{n=1}^{\Nev} \phi_n \phi_n^{\dag} \,,
\end{equation}
with $P$ the projector orthogonal to the space spanned by the $\Nev$ eigenvectors. 
In practice the stochastic sources used to estimate the disconnected contribution have support on the whole lattice and time dilution is not used there.  
To further reduce the numerical cost of the high-mode contribution to the quark loops, we have also implemented the all-mode averaging (AMA) technique~
\cite{Bali:2009hu,Blum:2012uh,Shintani:2014vja}. For an observable $\mathcal{O}$, we use the decomposition (exact in the infinite statistics limit)
\begin{equation}
\big\langle \mathcal{O} \big\rangle = \big\langle \mathcal{O}_{\rm sloppy} \big\rangle +  \big\langle \mathcal{O}_{\rm cor} \big\rangle \,,
\end{equation}
where 
\begin{equation}
\mathcal{O}_{\rm sloppy} = \frac{1}{N_1} \sum_{s=1}^{N_1}  \mathcal{O}^{(s)}_{\rm lp} \,, \quad 
\mathcal{O}_{\rm cor} = \frac{1}{N_2} \sum_{s=N_1+1}^{N_1+N_2}  \left( \mathcal{O}^{(s)}_{\rm hp} - \mathcal{O}^{(s)}_{\rm lp} \right) \,.
\end{equation}
In general, the estimator $\mathcal{O}$ requires the evaluation of quark propagators. For a given source ($s$), the estimator $\mathcal{O}^{(s)}_{\rm lp}$ is then defined by a fix number of deflated conjugate gradient iterations while $\mathcal{O}^{(s)}_{\rm hp}$ is obtained with exact (high precision) solves. In the correction term, the same stochastic noise is used for both low and high precision solves to maintain statistical correlations. 
The number of iterations defining the sloppy solves is tuned such that the correction term is small as compared to the statistical precision. 
In practice, we use 600 iterations at our finest lattice spacing while exact solves are defined by a residual smaller than $10^{-8}$. It is thus numerically beneficial to use $N_1 \gg N_2$. 
It appears that AMA leads to a modest gain for pseudoscalar loops. However, it is very efficient for the vector loops that require many more inversions of the Dirac operator. Although those vector loops are not used in the work presented here, they are needed for the calculation of the light pseudoscalar transition form factors presented in~\cite{Gerardin:2023naa}. 

\begin{figure}[t]
	\centering
	\includegraphics[width=0.49\textwidth]{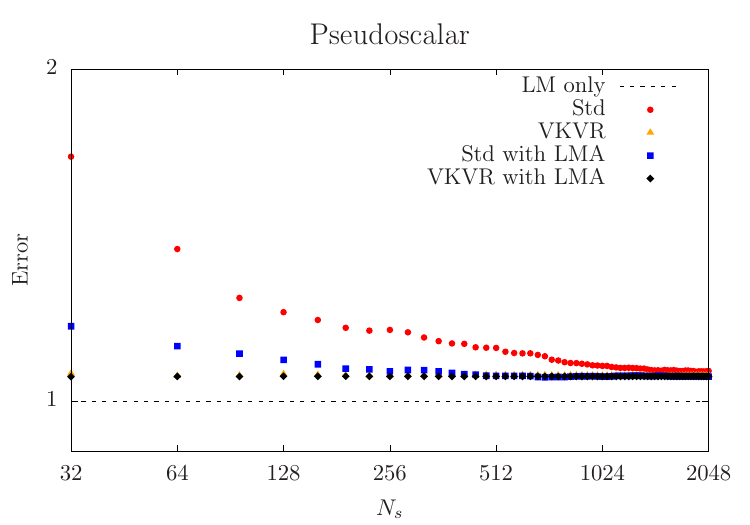}
	\includegraphics[width=0.49\textwidth]{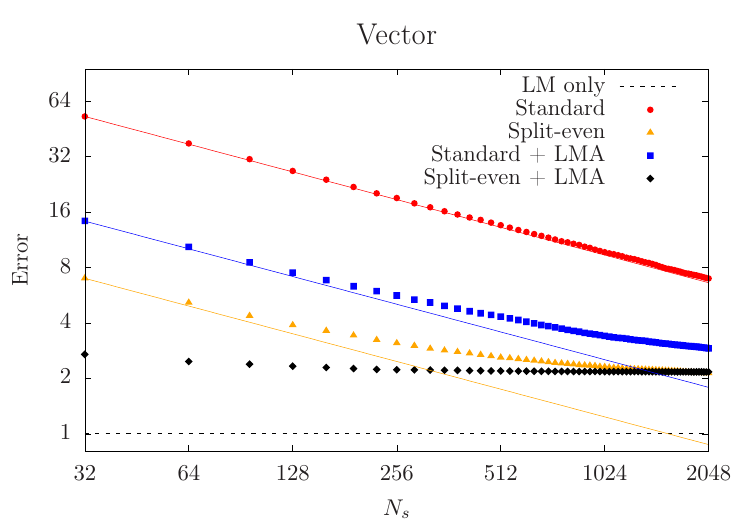}
	\caption{Comparison of the square root of the variance of the pseudoscalar (left) and vector (right) loops functions as a function of the number of stochastic sources $N_s$ for the different estimators described in the text. For the vector loop, we present the results for the ``light minus strange" flavor difference. The lines represent a perfect scaling of the variance assuming the variance is dominated by the stochastic noise. The horizontal dashed line is the error from the low-mode contribution only and is used as a normalization scale. }
	\label{fig:noise}
\end{figure}

Finally, when using the 4-link operator, we have implemented the Venkataraman-Kilcup variance reduction trick introduced in~\cite{Venkataraman:1997xi,Gregory:2007ev} to estimate the stochastic part of the disconnected contribution. 
Noting that both the 4-link operator $O_4$ and $(D^{\dag} D)^{-1}$ connects sites that are separated by an even number of gauge links, we obtain 
\begin{equation}
\mathrm{Tr} \left[ O_4 D^{-1} \right] = \mathrm{Tr} \left[ O_4 (D^{\dag }D)^{-1} D^{\dag} \right] 
\approx m_f \frac{1}{N_s}  \sum_{s}  \psi_s^{\dag} \, (O_4 \psi_s) \,, 
\end{equation}
with $\psi_s = D^{-1} \eta_s$, the solution vector for the source $(s)$. 
The last equality is valid only in the limit of infinite statistics and we note that the hopping term of the Dirac operator does not contribute such that the final estimator gets an explicit quark mass factor. An important feature of the VKVR trick is that it can be used for each flavor independently.

The various methods have been tested for both the pseudoscalar operator $O_4$ and for the conserved vector current 
\begin{equation}
J_{\mu}(x) = -\frac{1}{2} \eta_{\mu}(x) \left[ \chib(x+a\hat{\mu}) U_{\mu}^{\dag}(x) \chi(x) + \chib(x) U_{\mu}(x) \chi(x+a\hat{\mu}) \right] \,.
\label{eq:V}
\end{equation}
The comparison is performed at the level of the loop functions used to evaluate the disconnected diagrams. More precisely, we compute the quantities
\begin{subequations}
\label{eq:loops}
\begin{align}
L_4^{(f)}(t) &=  - \frac{1}{2} \left(\frac{a}{L}\right)^{3} \sum_{ \xv } \eta_4(x) \, \Tr \left[ \hat{O}_3 U_4(x) S_f(x+a\hat{t}, x) + U^{\dag}_{4}(x) \hat{O}_3 S_f(x, x+a\hat{t}) \right]  \,, \\
L_{V;\mu}^{(f)}(t) &= + \frac{1}{2} \left(\frac{a}{L}\right)^{3} \sum_{ \xv } \, \eta_{\mu}(x) \, \Tr \left[ S_f(x, x+a\hat{\mu})  U_{\mu}^{\dag}(x)  + S_f(x+a\hat{\mu}, x) U_{\mu}(x) \right]  \,,
\end{align}
\end{subequations}
where $S_f$ is the quark propagator with flavor $f$. The results are shown in \Fig{fig:noise}. For pseudoscalar loops, the VKVR trick is extremely efficient and performs better than LMA. With 32 sources, we already reach the gauge noise. For vector loops, the VKVR trick is not applicable and LMA leads to a significant gain. If one is only interested in the light minus strange flavor difference~\cite{Gerardin:2023naa}, as often happens with electromagnetic currents, then the split-even estimator introduced in~\cite{Giusti:2019kff} can be combined with LMA (see Appendix B of~\cite{Gerardin:2023naa}). With this improved estimator, one can see in~\Fig{fig:noise} that the gauge noise is reached for $N_s<100$ stochastic sources.

\section{Taste singlet pion and taste splitting \label{sec:pion}}

The staggered quark transformation reduces the number of tastes from 16 to 4 such that we are left with 16 pions that differ by tastes. In the continuum limit all tastes have equal mass but, at finite lattice spacing, taste interactions break this degeneracy. The Goldstone pion, with pseudoscalar taste ($\gamma_5 \otimes \gamma_5$) is the lightest pion and its mass has been tuned close to the physical pion mass in our simulations. 
Instead, the taste-singlet pion is the heaviest and its mass goes to the physical pion mass only in the continuum limit.

\subsection{Comparison of the 3-link and 4-link operators \label{sec:opcmp}}

\begin{figure}[t]
	\centering
	\includegraphics[width=0.29\textwidth]{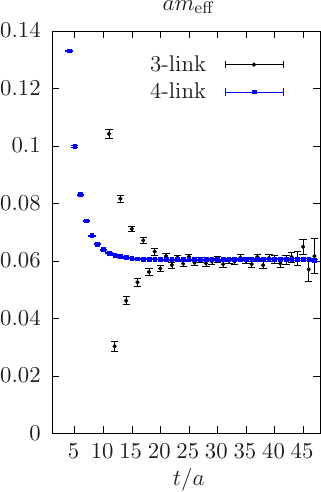} \hspace{2cm}
	\includegraphics[width=0.29\textwidth]{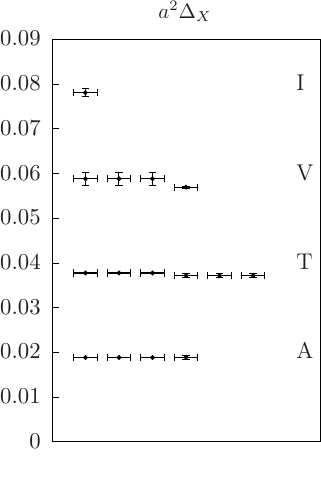}
	\caption{Left panel: effective mass plot for the 3-link and 4-link operators at our finest lattice spacing (note that a higher statistics is used for the 4-link operator). Right panel: mass splitting in the pion spectrum at our coarsest lattice spacing. The labels $X \in$ (P,A,T,V,I) refer to pseudoscalar, axial, tensor, vector and singlet tastes. The multiplet structure predicted by LO rS$\chi$PT is clearly observed.}
	\label{fig:O3O4}
\end{figure}

In this section we compare the 3-link and the 4-link operators, introduced in  \Section{sec:op}, for the taste-singlet pion where only the light quark-connected diagram contributes.
In general, a staggered correlation function projected on a given spatial momenta presents oscillations in time due to the contribution of the parity partner state that appears with an extra factor $(-1)^{t/a}$. Neglecting excited states, the staggered correlator can be described by 
\begin{equation}
C(t;\pv) = A \cosh\left( E_1(\pv) [T/2-t] \right) + (-1)^{t/a} B \cosh\left( E_2(\pv) [ T/2-t ] \right) \,,
\label{eq:cor}
\end{equation}
where $E_1$ and $E_2$ are the energies of the pion and its parity partner state, $T$ is the time extent of the lattice and $A$ and $B$ are some constant coefficients that depend on the details of the interpolating operator. Therefore, in the presence of an oscillating term, we define the smeared correlation function
\begin{equation} \label{eq:smr}
\widetilde{C}(t) = \frac{1}{4} C(t-a) + \frac{1}{2} C(t) + \frac{1}{4} C(t+a) \,,
\end{equation}
that projects (approximately) onto the pion state. With this definition, the suppression of the parity partner state is O$(a^2)$ and the effective mass can be defined by the usual logarithmic derivative of the smeared correlator.

In the left panel of \Fig{fig:O3O4}, we compare the effective mass for both the 3- and 4-link operators. In \cite{Altmeyer:1992dd} it was argued that the contribution of the parity partner state to the 4-link operator is strongly suppressed, a statement supported by our data  were no oscillations are visible at all. Thus, for the 4-link operator, the effective mass is obtained using the original correlator and the smearing in~\Eq{eq:smr} is only applied to the 3-link operator. 
As expected, both operators lead to compatible values of the pseudoscalar mass. The 4-link operator presents two advantages: first the plateau is reached at earlier times and does not exhibit oscillations, second, the VKVR trick described above is applicable for the disconnected contribution. For these reasons, only the 4-link operator is used in our analysis of the $\eta^{(\prime)}$ mesons described below. 

\subsection{Pion mass and taste-breaking effects \label{sec:taste}}

At our coarsest lattice spacing, we have computed the full pion spectrum and the result is shown on the right panel of \Fig{fig:O3O4}. 
We observe a near degeneracy for the axial-, tensor- and vector-tastes, as predicted by LO rS$\chi$PT~\cite{Lee:1999zxa,Bernard:2001yj,Aubin:2003mg}, and the mass splitting between the lightest taste-pseudoscalar and the heaviest taste-singlet pion is $\Delta \approx 300~$MeV. 
From rS$\chi$PT, sufficiently close to the continuum and chiral limits, the masses-squared of the different pion tastes are given by $(m_{\pi}^{X})^2 = \mu \left(m_u + m_d\right) + a^2 \Delta_X$ with $\mu$ a constant. The mass splitting contribution for each taste, labeled by $X \in (5,\mu5,\mu\nu,\mu,I)$, reads 
\begin{subequations}
\label{eq:chipt}
\begin{align}
	\Delta_{P} &= 0 \\
	\Delta_{A} &= \frac{16}{f^2}\left( C_1 + 3C_3 + C_4 + 3C_6 \right) \\
	\Delta_{T} &= \frac{16}{f^2}\left( 2C_3 + 2C_4 + 4C_6\right) \\
	\Delta_{V} &= \frac{16}{f^2}\left( C_1 + C_3 + 3C_4 + 3C_6 \right) \\
	\Delta_{I} &= \frac{16}{f^2}\left( 4C_3 + 4C_4 \right)\,,
\end{align}
\end{subequations}
with $f$ the pion decay constant in the chiral limit and $C_i$ are coefficients defined in Eq.~(13) of Ref.~\cite{Aubin:2003mg}. 
Our data is compatible with this pattern showing that SO(4) taste breaking terms are numerically small. The fact that the mass difference between each multiplet is roughly constant suggests that the coefficient $C_4$ is numerically dominant. This observation was already made in~\cite{Lee:1999zxa,Aubin:2003mg,Bernard:2001av,Aubin:2004wf,Aubin:2022hgm} for different discretizations of the action. 

\begin{figure}[t]
	\centering
	\includegraphics[width=0.59\textwidth]{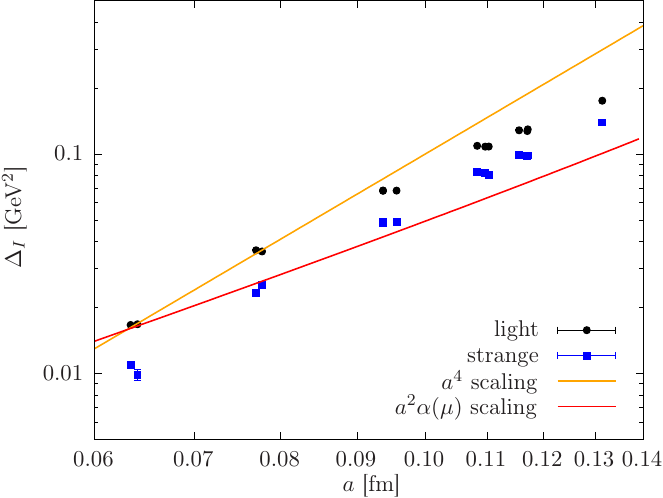}
	\caption{Taste-splitting $\Delta_I$ as a function of the lattice spacing. The orange line represents an exact $a^4$ scaling. Black circles (blue squares) correspond to the light (strange) contribution.}
	\label{fig:splitting}
\end{figure}

In \Fig{fig:splitting} we also show the taste-splitting $\Delta_I$ as a function of the lattice spacing. At leading order in rS$\chi$PT and in the strong coupling $\alpha_s$, one expects the mass splitting to scale approximately with $a^2 \alpha_s(\mu)$ where $\alpha_s$ is the strong running coupling in the $\overline{\rm MS}$ scheme evaluated at a scale $\mu \approx 1/a$.  As already pointed out in~\cite{Borsanyi:2020mff}, we observe that the taste-splitting decreases much faster than the leading $a^2 \alpha_s(\mu)$. 

Finally, although the mass splittings $\Delta_I$ for mesons composed of two light or two strange quarks are indeed close, we observe a small difference, depicted in~\Fig{fig:splitting}. For the taste-singlet $\eta_8$ meson, the prediction from rS$\chi$PT is~\cite{Bernard:2007qf}
\begin{equation}
m^{2}_{\eta_8,I} = \frac{1}{3} m^{2}_{uu,I} + \frac{2}{3} m^{2}_{ss,I}
\label{eq:EtaChiPT}
\end{equation}
and we expect strong taste-breaking effects for the $\eta$ meson. At our coarsest lattice spacing, $m_{\pi,I} \approx 440~$MeV while the taste-pseudoscalar pion is close to its physical value. Thus we expect discretization effects as large as $140$~MeV for the $\eta$ meson mass at our coarsest lattice spacing. 
 
\section{Computation of the masses and amplitudes  of the $\eta$ and $\etap$ mesons} \label{sec:res}

The correlation matrix in \Eq{eq:C2iso} quickly becomes noisy due to the presence of disconnected contributions and it is difficult to find a time interval where the signal is clearly dominated by the ground state while keeping statistical errors under control. 
In~\cite{Michael:2013gka,Ottnad:2015hva}, it was observed that excited states mostly contribute to the connected contribution, that can be computed with much higher statistical precision than the disconnected diagrams. We thus remove the excited state contributions in the connected diagrams using the following strategy. 
At large time, where excited-state contributions are negligible, the connected correlators $\mathcal{C}_{l}(t)$ and $\mathcal{C}_{s}(t)$ in \Eq{eq:C2iso} are well-described by a single exponential. In a second step, $\mathcal{C}_{l}(t)$ and $\mathcal{C}_{s}(t)$ are replaced by their ground state contribution at times $t>0$. If the assumption that excited states do not contribute significantly to the disconnected part is correct, we thus expect plateaus at very early times. The validity of this approximation can be tested a posteriori. 
However, in contrast to the standard effective mass, computed from the unsubtracted correlator, the plateau is not guaranteed to be approached from above. 

\subsection{Extraction of masses and overlaps}

In this section, we discuss the two different strategies that have been used to extract  the masses and overlap factors of the $\eta$ and $\etap$ mesons. 

\subsubsection{Masses and amplitudes from the GEVP} \label{sec:gevp}

\begin{figure}[t]
	\centering
	\includegraphics[width=0.29\textwidth]{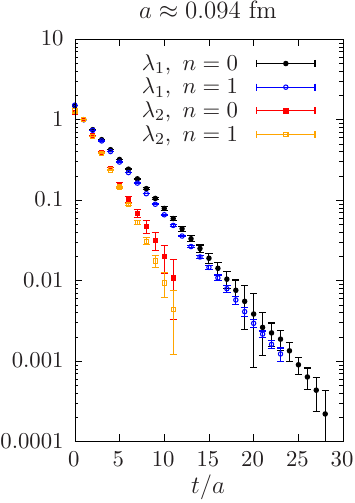} \quad
	\includegraphics[width=0.29\textwidth]{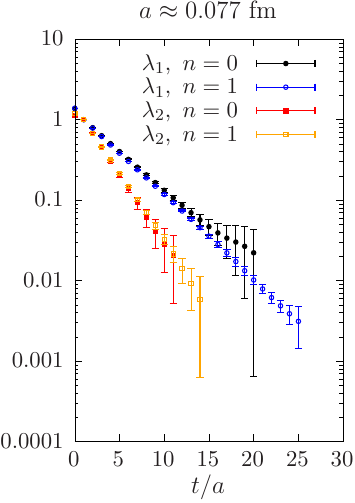} \quad
	\includegraphics[width=0.29\textwidth]{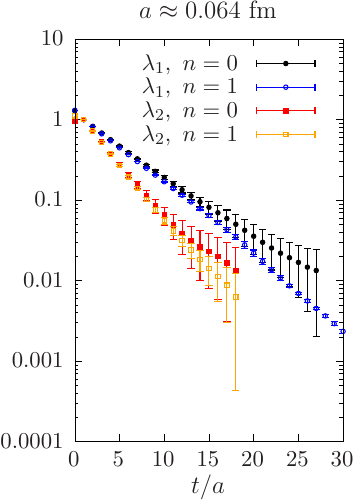} 
	\caption{Time dependence of the eigenvalues associated to the $\eta$ ($\lambda_1$) and $\etap$ ($\lambda_2$) mesons both at vanishing ($n=0$) and non-vanishing ($n=1$) momenta where $|\vec{p}| = n(2\pi/L)$. From left to right, our three finest lattice spacings.}
	\label{fig:eigenvalues}
\end{figure}

A well established method to extract information about excited states is to solve the Generalized Eigenvalue Problem (GEVP)~\cite{Blossier:2009kd}
\begin{equation}
C^{\rm 2pt}(t) v_n(t,t_0) = \lambda_n(t,t_0) C^{\rm 2pt}(t_0) v_n(t,t_0) \,,
\label{eq:gevp}
\end{equation}
where $\lambda_n$ and $v_n$ are the eigenvalues and eigenvectors and $t_0$ is a free parameter. 
The eigenvectors are normalized such that
\begin{equation}
v_m(t,t_0)^{\rm T} \, C^{\rm 2pt}(t_0) \, v_n(t,t_0) = \delta_{n,m} \,.
\label{eq:normeigen}
\end{equation}
In practice, this equation is solved independently for each momentum and the dependence on the momentum is not shown explicitly. 
In~\Section{sec:methodology}, it was argued that using $C^{\rm 2pt} = \overline{C}$ at vanishing momentum removes possible constant contributions due a partial sampling of the topological charge. 
In addition, we observe that using $C^{\rm 2pt} = \overline{C}$ leads to smaller statistical error for $\vec{p} = \vec{0}$. Thus, our preferred strategy is to use the combination with $C^{\rm 2pt} = \overline{C}$ at zero momentum and $C^{\rm 2pt} = C$ at non-vanishing momentum. 
The time dependence of the eigenvalues for a few ensembles is shown in \Fig{fig:eigenvalues}.
We observe that the eigenvalues at non-vanishing momentum tend to have smaller statistical uncertainties. 
During the analysis, we did not see a significant impact on the choice of $t_0$. As large values lead to noisy results, we simply set $t_0=a$ in the following. 

\begin{figure}[t!]
	\centering
	\includegraphics[width=0.42\textwidth]{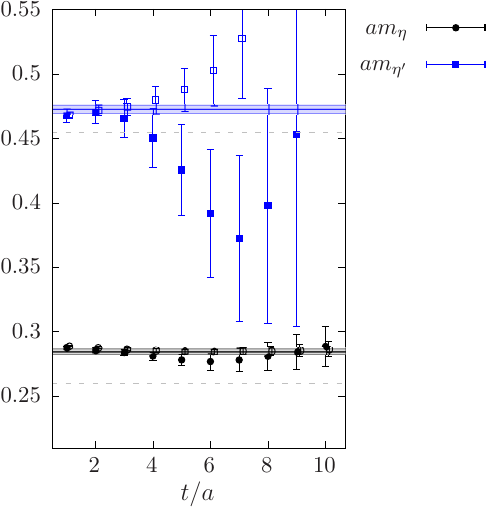} \quad
	\includegraphics[width=0.42\textwidth]{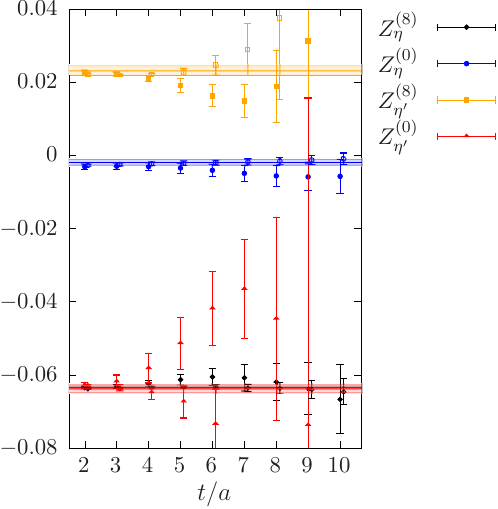}
	\vspace{-0.05cm}
	\caption{Left: Effective masses at vanishing momentum (filled symbols) and at non-vanishing momentum (open symbols). At non-vanishing momentum, the energies are shifted assuming the continuum relativistic dispersion relation. The bands correspond to our best estimate using a combined fit to both eigenvalues. Right: similar plot for the four overlap factors. The results are given at $a \approx 0.094~$fm.}
	\label{fig:gevp1}
\end{figure}

The largest eigenvalue is associated with the ground state $\eta$ meson. A simple estimator of the pseudoscalar meson mass is given by the logarithmic derivative~\cite{Blossier:2009kd} 
\begin{equation}
aE_n^{\rm eff}(t) = \log \frac{ \lambda_n(t,t_0) }{ \lambda_n(t+a,t_0) } \,, 
\label{eq:gevpE}
\end{equation}
with $n=\eta,\etap$. 
In this equation, the contribution from backward propagating mesons, which is numerically small compared to the statistical error, is neglected. 
At sufficiently large time, the effective mass can eventually be fitted to a constant. For the $\etap$, we could expect sizable excited states contributions. However, our data suggests a plateau at early times and we are not able to resolve these excited states given our statistical precision (see \Fig{fig:gevp1}).
The overlap factors can be extracted from the eigenvectors through~\cite{Blossier:2009kd} 
\begin{equation} \label{eq:gevpZ1}
Z_n^{(i) {\rm eff} }(t) = \sqrt{2 E_n} \left( \frac{\lambda_n(t,t_0)}{ \lambda_n(t+a,t_0) } \right)^{ (t - t_0/2)/a}  \ \sum_{j} C_{ij}(t) \, v_{nj}(t,t_0)  \,,
\end{equation}
where $E_n$ is the energy of eigenstate $n$ extracted using~\Eq{eq:gevpE}. 
If the subtracted correlation function $\overline{C}$ is used, the effective overlap factor reads
\begin{equation} \label{eq:gevpZ2}
Z_n^{(i) {\rm eff} }(t) = \sqrt{2 E_n} \left( 2 \sinh\left( \frac{E_n \Delta t}{2} \right) \right)^{-1/2}  \left( \frac{\lambda_n(t,t_0)}{ \lambda_n(t+a,t_0) } \right)^{(t + \Delta t/4 - t_0/2)/a}  \ \sum_{j} \overline{C}_{ij}(t) \, v_{nj}(t,t_0)  \,. 
\end{equation}            
and again, plateaus seem to be reached at early times. 
An appealing feature of this method is the possibility to plot the effective mass or overlap factors as a function of $t$, making it easier to find a reasonable fit interval. However, these estimators tend to be noisy and the plateaus are lost at early times, especially for the $\etap$ meson, as can be seen in \Fig{fig:gevp1}. 

\begin{figure}[t]
	\centering
	\includegraphics[width=0.275\textwidth]{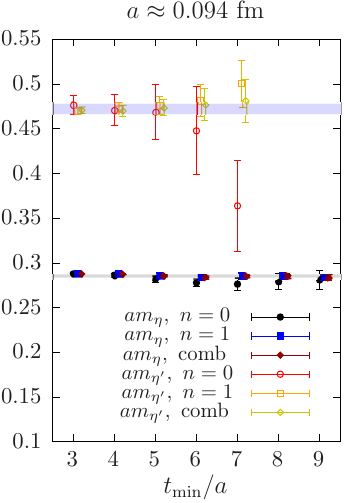} \hspace{0.5cm}
	\includegraphics[width=0.275\textwidth]{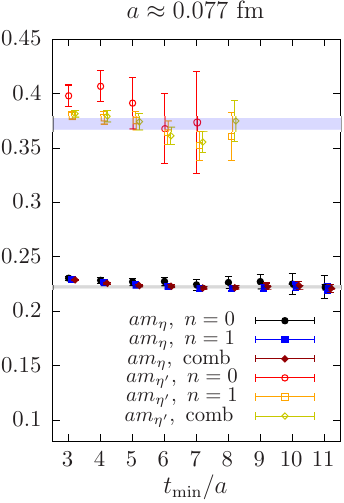} \hspace{0.5cm}
	\includegraphics[width=0.28\textwidth]{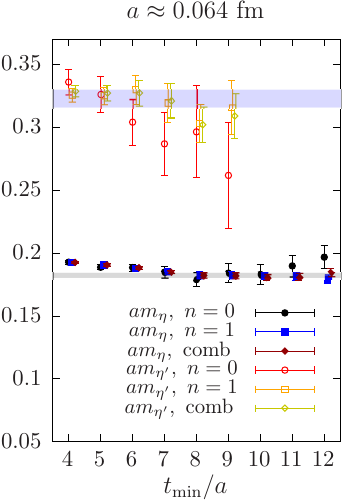}
	\vspace{-0.05cm}
	\caption{Masses of the $\eta$ and $\etap$ mesons extracted from a fit to the eigenvalues as a function of $t_{\rm min}$. For each state we either fit data for $n=0$, $n=1$ or both momenta (comb). The bands correspond to our best estimates using the correlator fit method.} 
	\label{fig:gevp2}
\end{figure}

If one is only interested in the extraction of the meson masses, it is possible to fit the eigenvalues. In that case, we can benefit from having different values of the momenta by performing combined fits and assuming the validity of the continuum dispersion relation. We find that this method leads to better behaved fits. 
In~\Fig{fig:gevp2}, we present the results of our correlated fits as a function of $t_{\rm min}$, the first time-slice included in the fit and which is chosen to be the same for both momenta. The final value is obtained once a plateau in $t_{\rm min}$ is observed. The $\chi^2/\dof$ of all ensembles are in the range $[0.6:1.9]$ for the $\eta$ meson and $[0.4:1.7]$ for the $\etap$ meson. 
For comparison, we also present the results obtained by fitting the eigenvalues at a single momentum: within uncertainties, we do not observe any deviation from the continuum dispersion relation and the combined fits tend to be more stables. 

\subsubsection{Fits of the correlation matrix}

\begin{figure}[t!]
	\centering
	\includegraphics[width=0.275\textwidth]{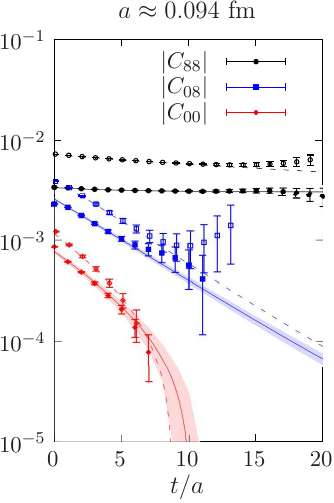} \hspace{0.5cm}
	\includegraphics[width=0.275\textwidth]{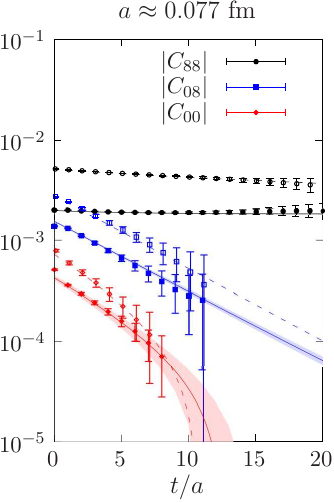} \hspace{0.5cm}
	\includegraphics[width=0.275\textwidth]{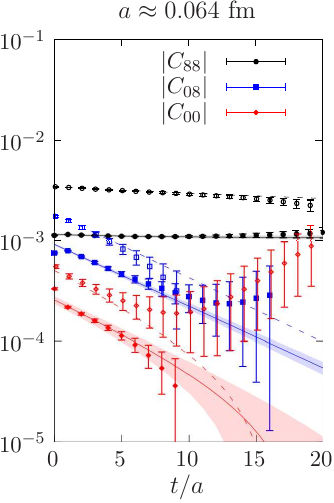}
	\caption{The bands represents the results of a fit to the six correlation functions in \Eq{eq:C2iso} at our three finest lattice spacings as described in the main text. In these fits, $C^{\rm 2pt} = \overline{C}$ and $C^{\rm 2pt} = C$ are used at vanishing (plain symbols) and non-vanishing (open symbols) momenta respectively. For clarity, the correlation functions have been rescaled by a factor $\exp(-\gamma t)$ with some arbitrary $\gamma$.}
	\label{fig:corfit}
\end{figure}

Instead of solving the GEVP, it is also possible to perform direct fits of the correlation matrix. In this case, we benefit from having two momenta in the extraction of  the masses and the overlap factors. 
Based on \Eq{eq:spectraldec}, we assume the following fit ansatz, with six independent parameters,
\begin{equation}
C_{ij}(t;\pv) =  \frac{Z_{\eta}^{(i)} Z_{\eta}^{(j)} }{2 E_{\eta}(\pv) } e^{-E_{\eta}(\pv) \, t} + \frac{Z_{\etap}^{(i)} Z_{\etap}^{(j)} }{ 2 E_{\etap}(\pv) }  e^{-E_{\etap}(\pv) \, t} \,,
\label{eq:specdec}
\end{equation}
that is expected to provide a good description of our data if higher excited states can be neglected.  
Assuming the validity of the dispersion relation, it is actually possible to extend the fit to both values of the pseudoscalar momenta without including any new fit parameter. If one considers the subtracted correlation function $\overline{C}(t)$, the fit ansatz becomes
\begin{equation}
\overline{C}_{ij}(t,\pv) =  \frac{ Z_{\eta}^{(i)} Z_{\eta}^{(j)} }{ 2 E_{\eta}(\pv) } \left( 1 - e^{-E_{\eta}(\pv)  \Delta t} \right) e^{-E_{\eta}(\pv) \,  t} + \frac{Z_{\etap}^{(i)} Z_{\etap}^{(j)} }{2 E_{\etap}(\pv) }  \left( 1 - e^{-E_{\etap}(\pv) \Delta t} \right) e^{-E_{\etap}(\pv) \, t}  \,. 
\end{equation}
As before, our preferred fit strategy is to use the subtracted correlator $\overline{C}$ at vanishing momentum and the original correlator $C$ at non-vanishing momentum. This is based on the observation that $\overline{C}(t)$ is statistically more precise as compared to $C(t)$ at vanishing momentum. 
Other combinations have been considered but will only serve as cross-checks in the final continuum extrapolation. Examples of fits are depicted in~\Fig{fig:corfit}.

With two momenta, we have to fit six correlators simultaneously. To avoid bias due to by-eye selection and to simplify the fit-range selection, we follow the strategy proposed in~\cite{Jay:2020jkz} and use the Akaike information criterion to select the best fit ranges. In practice, we vary the initial time-slice for each of the six correlators and we associate a weight that depends on the correlated chi-square and the number of degrees of freedom. Explicitly the relative weights are
\begin{equation}
w_i \propto \exp\left( -\frac{1}{2} \mathrm{AIC} \right) \,,\quad \mathrm{AIC} = \chi^2 - n_{\rm data} \,,
\end{equation}
where $n_{\rm data}$ is the total number of point included in the fit. 
On each ensemble, our final result is obtained by a weighted average over all these fits. In practice we observe that the highest weights have typical chi-squared per degree of freedom in the range $\chi^2/\ndof \in [0.7:1.6]$. 

\begin{figure}[t]
	\centering
	\includegraphics[width=0.75\textwidth]{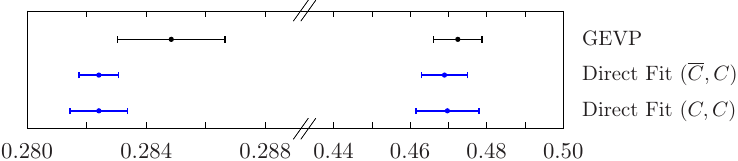} \\[5mm]
	\includegraphics[width=0.75\textwidth]{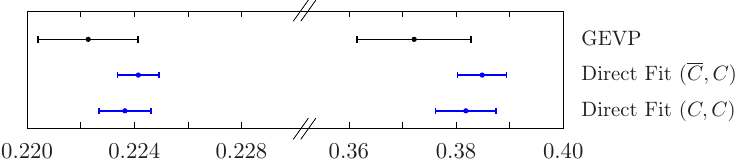} \\[5mm]
	\includegraphics[width=0.75\textwidth]{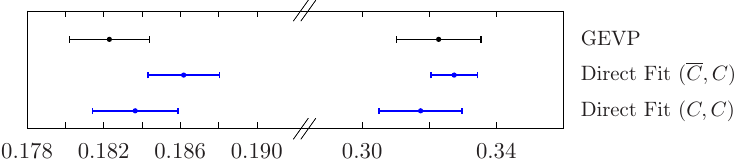} 
	\caption{Masses of the $\eta$ (left) and $\etap$ (right) mesons, in lattice units, using the different methods presented in the main text. The black point is obtained from a fit to the eigenvalues. The blue points are obtained from fits of the correlation matrix. The notation $(A,B)$ refers to the correlator ($C$ or $\overline{C}$) used for $|p|=0$ and $|p| \neq 0$ respectively. From top to bottom, our three finest lattice spacings.}
	\label{fig:cmp}
\end{figure}

In general, we observe that these correlator fits lead to slightly more precise results. 
A direct comparison of the two different methods analyzed in this work is given in \Fig{fig:cmp} for our three finest lattice spacings.

\subsection{Extrapolation to the continuum limit}  \label{sec:extrap}

To extrapolate our results to the continuum limit, we assume the ansatz 
\begin{equation}
m_X(a,X_l,X_s) = m_X^{\rm phys} + \beta_2 (\Lambda a)^2 + \beta_3 \alpha_s^n(a^{-1}) a^2 + \beta_4 (\Lambda a)^4 + \gamma_l X_l + \gamma_s X_s
\end{equation}
where $n=2$ or 3 and $\Lambda = 0.5~$GeV is a typical QCD scale. Here,  $\alpha_s(a^{-1})$ is the strong coupling constant in the $\overline{\rm MS}$ scheme evaluated at the scale $1/a$. The quantities $X_l$ and $X_s$ are proxies for the slight mistuning of the light and strange quark masses 
\begin{equation}
X_l = \frac{M_{\pi^0}^2}{ ( M_{\pi^0}^{\rm phys} )^2} - 1 \,, \quad X_s= \frac{M_{K_{\chi}}^2}{ ( M_{K_{\chi}}^{\rm phys} )^2} - 1
\end{equation}
and $M_{K_{\chi}}^2 = \frac{1}{2} \left( M_{K_0}^2 + M_{K_+}^2 - M_{\pi^+}^2 \right)$. See~\cite{Borsanyi:2020mff} for more details. 

\begin{figure}[t!]
	\centering
	\includegraphics[width=0.3\textwidth]{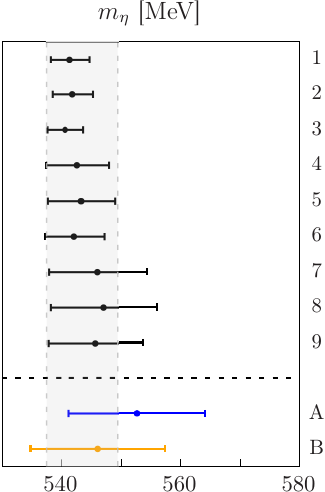} \hspace{2cm}
	\includegraphics[width=0.315\textwidth]{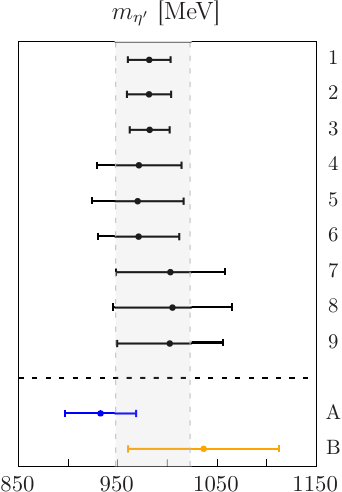}

	\caption{The nine variations used to extrapolate our data to the continuum limit. The labels are explained in the main text. The gray bands represent our  total uncertainties. The variation A is the result of our continuum extrapolation when using the GEVP method while the variation B corresponds to the case where the original correlator is used for vanishing momentum. }
	\label{fig:syst}
\end{figure}

In our fits, the correlations between the independent variables $a$, $X_l$ and $X_s$ are not taken into account. Since the statistical uncertainty associated with the continuum limit is estimated using the jackknife procedure this approximation only prevents us from giving a clear meaning to the quality of the fit. 
Our set of 13 ensembles, spread among 6 lattice spacings, does not allow to fit $\beta_3$ and $\beta_4$ simultaneously. We thus estimate the systematic uncertainty by looking at variations keeping either $\beta_3$ or $\beta_4$ and adding cuts in the lattice spacing. In total, we perform 9 analyses :
\begin{itemize}
\item analysis (1) : $\beta_3 = 0$, no cut in the lattice spacing
\item analysis (2) : $\beta_4 = 0$, $n=2$, no cut in the lattice spacing
\item analysis (3) : $\beta_4 = 0$, $n=3$, no cut in the lattice spacing
\item analysis (4) : $\beta_3 = 0$, cut at $a = 0.125~$fm
\item analysis (5) : $\beta_4 = 0$, $n=2$, cut at $a = 0.125~$fm
\item analysis (6) : $\beta_4 = 0$, $n=3$, cut at $a = 0.125~$fm
\item analysis (7) : $\beta_3 = 0$, cut at $a = 0.113~$fm
\item analysis (8) : $\beta_4 = 0$, $n=2$, cut at $a = 0.113~$fm
\item analysis (9) : $\beta_4 = 0$, $n=3$, cut at $a = 0.113~$fm
\end{itemize}
and the systematic error is estimated by computing the root-mean-squared deviation of the fit results compared to the flat average. The results obtained for each variation are summarized in \Fig{fig:syst}. In addition, we show the results of two additional variations called A and B. The result A is obtained using the GEVP method described in \Section{sec:gevp}. The analysis B is the same as the main analysis but using the original correlator $C$ instead of $\overline{C}$ at vanishing momentum. The results agree with the main analysis although with larger statistical errors.

\begin{figure}[t!]
	\centering
	\includegraphics[width=0.32\textwidth]{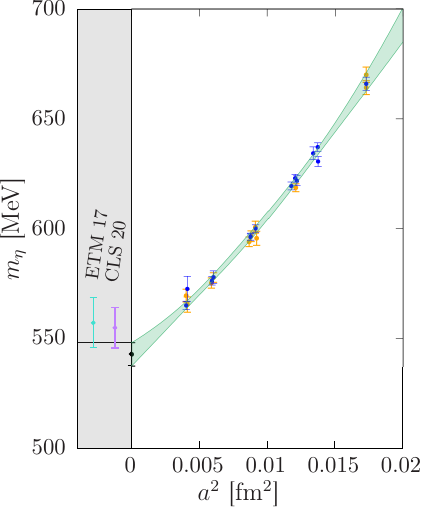}
	\includegraphics[width=0.32\textwidth]{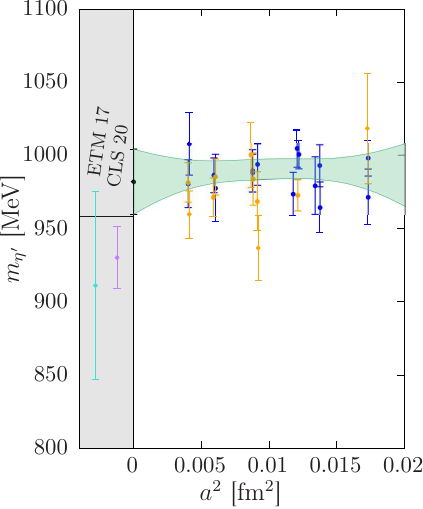}
	\includegraphics[width=0.313\textwidth]{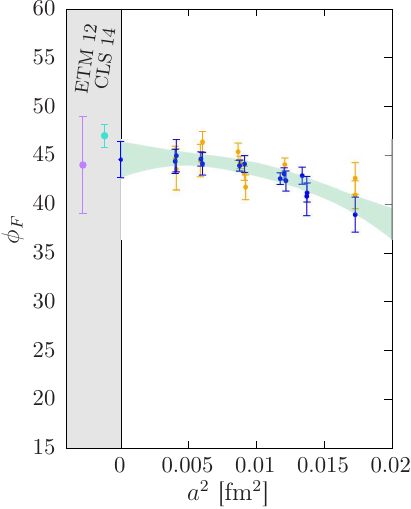}
	
	\caption{Continuum extrapolation of the $\eta$ meson mass (left), $\etap$ meson mass (middle) and of the mixing angle $\phi_F$ (right). The blue circles represent the lattice data obtained using large-volume simulation with $L=6~\fm$. The yellow square symbols represents the lattice data obtained using smaller volumes with $L=3~\fm$. Only the large volume ensembles are included in the continuum extrapolation.}
	\label{fig:extrap}
\end{figure}

Typical continuum extrapolations, for both the $\eta$ and the $\etap$, are shown in~\Fig{fig:extrap}. 
For the $\eta$ meson, we observe a strong dependence on the lattice spacing, as anticipated in Section~\ref{sec:taste} where we studied the taste-splitting as a function of the lattice spacing. 
In those fits, we focus on the large volume ensembles with $L \geq 6~$fm and the data-points on smaller physical volumes are only shown for completeness as they are used in our recent work on the pseudoscalar transition form factors~\cite{Gerardin:2023naa}. They also indicate that finite-volume effects are small and can be neglected at our level of precision. 
For the $\etap$ pseudoscalar meson, we observe a very mild dependence on the lattice spacing. Again, the results obtained using smaller volumes suggest that finite-volume effects are under control. 

Finally, we also present our continuum extrapolation of the mixing angle defined in terms of amplitudes in the flavor basis. The overlap factors play an important role in the extraction of form factors~\cite{Gerardin:2023naa}.

\section{Conclusion} 
\label{sec:ccl}

In this paper, we have presented a lattice calculation of the $\eta$ and $\etap$ masses directly at the physical pion mass, using rooted staggered quarks. 
We have investigated different taste-singlet operators 
and various noise-reduction methods to compute the quark-disconnected contribution efficiently. The pseudoscalar taste-singlet 4-link operator essentially suppress the staggered oscillations such that the data analysis is greatly simplified. In addition, the VKVR trick~\cite{Venkataraman:1997xi,Gregory:2007ev} can be used to compute the quark-disconnected contribution efficiently. 
We have computed the masses and the mixing angles (in terms of amplitudes). Those quantities are relevant for the calculation of the pseudoscalar transition form factors that are a key input in the estimate of the pseudoscalar-pole contribution to the hadronic light-by-light scattering in the muon $(g-2)$~\cite{Gerardin:2023naa}. For the $\eta$ meson, we observe large taste-breaking effects that are due to the use of taste-singlet operators. On the other hand, we observe a very mild continuum extrapolation for the $\etap$ meson mass. 
After extrapolating our data to the continuum limit, we obtain
\begin{subequations}
\begin{align}
m_{\eta} &= 543.5 \pm 5.2 \pm 2.2_{\rm cont} \ [5.6] ~\MeV \,, \\
m_{\etap} &= 986 \pm 35 \pm 14_{\rm cont} \ [38]~\MeV \,,
\end{align}
\end{subequations}
where the first error is statistical and the second is the systematic uncertainty associated with the continuum extrapolation. The total uncertainty is given in   brackets. Those values are compatible with the experimental values $m_{\eta} = 547.862(17)~\MeV$, $m_{\eta^{\prime}} = 957.78(06)~\MeV$~\cite{ParticleDataGroup:2024cfk}. 
A recent lattice calculation based on  $N_f = 2+1$ O($a$)-improved Wilson-clover ensemble quotes $m_{\eta} = 554.7(9.2)$~MeV and $m_{\etap} = 930(21)$~MeV~\cite{Bali:2021qem}. Using $N_f = 2+1+1$ twisted-mass fermions, the authors of~\cite{Ottnad:2017bjt} found $m_{\eta} = 557(11)$~MeV and $m_{\etap} = 911(64)$~MeV. 
Our calculation strongly supports the validity of rooting in the staggered quark formalism. 
In the future, it would be interesting to extend our study to include decay constants of the singlet mesons~\cite{Ottnad:2017bjt,Bali:2021qem}. 
They play an important role in the short distance behavior of the pseudoscalar transition form factors~\cite{Bali:2021qem,Gerardin:2023naa}, but also in the description of the partial decay widths of pseudoscalar mesons to two photons~\cite{Gerardin:2019vio,Gerardin:2023naa,ExtendedTwistedMass:2023jan,Alexandrou:2023lia}.

Regarding the importance of the staggered quarks discretization in precision tests of the Standard Model of particle physics, increasing the precision to provide stronger tests of the rooting procedure is certainly worthwhile. 
To further improve these results, a larger basis of interpolating operators for the GEVP should enhance the excited state extraction. This could be done using gaussian-smeared correlators or using a combination of the 4-link and 3-link operators that have different couplings to excited states. Also important would be to increase the number of measurements to check for possible long-range autocorrelations, especially in the extraction of the $\etap$ meson mass at fine lattice spacings. 


\begin{acknowledgments}
We thank all the members of the Budapest-Marseille-Wuppertal collaboration for helpful discussions and the access to the gauge ensembles used in this work. 
This publication received funding from the Excellence Initiative of Aix-Marseille University - A*Midex, a French ``Investissements d'Avenir" programme, AMX-18-ACE-005 and from the French National Research Agency under the contract ANR-20-CE31-0016. 
The computations were performed on Joliot-Curie at CEA's TGCC, on Jean Zay at IDRIS, on SuperMUC-NG at Leibniz Supercomputing Centre in M\"unchen, on HAWK at the High Performance Computing Center in Stuttgart and on JUWELS at Forschungszentrum J\"ulich. We thank GENCI (grants A0080511504, A0100511504 and A0120511504) and the Gauss Centre for Supercomputing (projects pn73xi and wprecision) for awarding us computer time on these machines.
Centre de Calcul Intensif d'Aix-Marseille (CCIAM) is acknowledged for granting access to its high performance computing resources.
\end{acknowledgments}


\bibliography{biblio}{}

\begin{thebibliography}{57}%
\makeatletter
\providecommand \@ifxundefined [1]{%
 \@ifx{#1\undefined}
}%
\providecommand \@ifnum [1]{%
 \ifnum #1\expandafter \@firstoftwo
 \else \expandafter \@secondoftwo
 \fi
}%
\providecommand \@ifx [1]{%
 \ifx #1\expandafter \@firstoftwo
 \else \expandafter \@secondoftwo
 \fi
}%
\providecommand \natexlab [1]{#1}%
\providecommand \enquote  [1]{``#1''}%
\providecommand \bibnamefont  [1]{#1}%
\providecommand \bibfnamefont [1]{#1}%
\providecommand \citenamefont [1]{#1}%
\providecommand \href@noop [0]{\@secondoftwo}%
\providecommand \href [0]{\begingroup \@sanitize@url \@href}%
\providecommand \@href[1]{\@@startlink{#1}\@@href}%
\providecommand \@@href[1]{\endgroup#1\@@endlink}%
\providecommand \@sanitize@url [0]{\catcode `\\12\catcode `\$12\catcode
  `\&12\catcode `\#12\catcode `\^12\catcode `\_12\catcode `\%12\relax}%
\providecommand \@@startlink[1]{}%
\providecommand \@@endlink[0]{}%
\providecommand \url  [0]{\begingroup\@sanitize@url \@url }%
\providecommand \@url [1]{\endgroup\@href {#1}{\urlprefix }}%
\providecommand \urlprefix  [0]{URL }%
\providecommand \Eprint [0]{\href }%
\providecommand \doibase [0]{https://doi.org/}%
\providecommand \selectlanguage [0]{\@gobble}%
\providecommand \bibinfo  [0]{\@secondoftwo}%
\providecommand \bibfield  [0]{\@secondoftwo}%
\providecommand \translation [1]{[#1]}%
\providecommand \BibitemOpen [0]{}%
\providecommand \bibitemStop [0]{}%
\providecommand \bibitemNoStop [0]{.\EOS\space}%
\providecommand \EOS [0]{\spacefactor3000\relax}%
\providecommand \BibitemShut  [1]{\csname bibitem#1\endcsname}%
\let\auto@bib@innerbib\@empty
\bibitem [{\citenamefont {Adler}(1969)}]{Adler:1969gk}%
  \BibitemOpen
  \bibfield  {author} {\bibinfo {author} {\bibfnamefont {S.~L.}\ \bibnamefont
  {Adler}},\ }\bibfield  {title} {\bibinfo {title} {{Axial vector vertex in
  spinor electrodynamics}},\ }\href {https://doi.org/10.1103/PhysRev.177.2426}
  {\bibfield  {journal} {\bibinfo  {journal} {Phys. Rev.}\ }\textbf {\bibinfo
  {volume} {177}},\ \bibinfo {pages} {2426} (\bibinfo {year}
  {1969})}\BibitemShut {NoStop}%
\bibitem [{\citenamefont {Bell}\ and\ \citenamefont
  {Jackiw}(1969)}]{Bell:1969ts}%
  \BibitemOpen
  \bibfield  {author} {\bibinfo {author} {\bibfnamefont {J.~S.}\ \bibnamefont
  {Bell}}\ and\ \bibinfo {author} {\bibfnamefont {R.}~\bibnamefont {Jackiw}},\
  }\bibfield  {title} {\bibinfo {title} {{A PCAC puzzle: $\pi^0 \to \gamma
  \gamma$ in the $\sigma$ model}},\ }\href {https://doi.org/10.1007/BF02823296}
  {\bibfield  {journal} {\bibinfo  {journal} {Nuovo Cim. A}\ }\textbf {\bibinfo
  {volume} {60}},\ \bibinfo {pages} {47} (\bibinfo {year} {1969})}\BibitemShut
  {NoStop}%
\bibitem [{\citenamefont {Belavin}\ \emph {et~al.}(1975)\citenamefont
  {Belavin}, \citenamefont {Polyakov}, \citenamefont {Schwartz},\ and\
  \citenamefont {Tyupkin}}]{Belavin:1975fg}%
  \BibitemOpen
  \bibfield  {author} {\bibinfo {author} {\bibfnamefont {A.~A.}\ \bibnamefont
  {Belavin}}, \bibinfo {author} {\bibfnamefont {A.~M.}\ \bibnamefont
  {Polyakov}}, \bibinfo {author} {\bibfnamefont {A.~S.}\ \bibnamefont
  {Schwartz}},\ and\ \bibinfo {author} {\bibfnamefont {Y.~S.}\ \bibnamefont
  {Tyupkin}},\ }\bibfield  {title} {\bibinfo {title} {{Pseudoparticle Solutions
  of the Yang-Mills Equations}},\ }\href
  {https://doi.org/10.1016/0370-2693(75)90163-X} {\bibfield  {journal}
  {\bibinfo  {journal} {Phys. Lett. B}\ }\textbf {\bibinfo {volume} {59}},\
  \bibinfo {pages} {85} (\bibinfo {year} {1975})}\BibitemShut {NoStop}%
\bibitem [{\citenamefont {'t~Hooft}(1976)}]{tHooft:1976rip}%
  \BibitemOpen
  \bibfield  {author} {\bibinfo {author} {\bibfnamefont {G.}~\bibnamefont
  {'t~Hooft}},\ }\bibfield  {title} {\bibinfo {title} {{Symmetry Breaking
  Through Bell-Jackiw Anomalies}},\ }\href
  {https://doi.org/10.1103/PhysRevLett.37.8} {\bibfield  {journal} {\bibinfo
  {journal} {Phys. Rev. Lett.}\ }\textbf {\bibinfo {volume} {37}},\ \bibinfo
  {pages} {8} (\bibinfo {year} {1976})}\BibitemShut {NoStop}%
\bibitem [{\citenamefont {Chen}\ \emph {et~al.}(2023)\citenamefont {Chen},
  \citenamefont {Jiang}, \citenamefont {Chen}, \citenamefont {Liu},
  \citenamefont {Sun},\ and\ \citenamefont {Yang}}]{Chen:2021dvn}%
  \BibitemOpen
  \bibfield  {author} {\bibinfo {author} {\bibfnamefont {F.}~\bibnamefont
  {Chen}}, \bibinfo {author} {\bibfnamefont {X.}~\bibnamefont {Jiang}},
  \bibinfo {author} {\bibfnamefont {Y.}~\bibnamefont {Chen}}, \bibinfo {author}
  {\bibfnamefont {K.-F.}\ \bibnamefont {Liu}}, \bibinfo {author} {\bibfnamefont
  {W.}~\bibnamefont {Sun}},\ and\ \bibinfo {author} {\bibfnamefont {Y.-B.}\
  \bibnamefont {Yang}},\ }\bibfield  {title} {\bibinfo {title} {{Glueballs at
  physical pion mass*}},\ }\href {https://doi.org/10.1088/1674-1137/accc1c}
  {\bibfield  {journal} {\bibinfo  {journal} {Chin. Phys. C}\ }\textbf
  {\bibinfo {volume} {47}},\ \bibinfo {pages} {063108} (\bibinfo {year}
  {2023})},\ \Eprint {https://arxiv.org/abs/2111.11929} {arXiv:2111.11929
  [hep-lat]} \BibitemShut {NoStop}%
\bibitem [{\citenamefont {Vadacchino}(2023)}]{Vadacchino:2023vnc}%
  \BibitemOpen
  \bibfield  {author} {\bibinfo {author} {\bibfnamefont {D.}~\bibnamefont
  {Vadacchino}},\ }\bibfield  {title} {\bibinfo {title} {{A review on Glueball
  hunting}},\ }in\ \href@noop {} {\emph {\bibinfo {booktitle} {{39th
  International Symposium on Lattice Field Theory}}}}\ (\bibinfo {year}
  {2023})\ \Eprint {https://arxiv.org/abs/2305.04869} {arXiv:2305.04869
  [hep-lat]} \BibitemShut {NoStop}%
\bibitem [{\citenamefont {Witten}(1979)}]{Witten:1979vv}%
  \BibitemOpen
  \bibfield  {author} {\bibinfo {author} {\bibfnamefont {E.}~\bibnamefont
  {Witten}},\ }\bibfield  {title} {\bibinfo {title} {{Current Algebra Theorems
  for the U(1) Goldstone Boson}},\ }\href
  {https://doi.org/10.1016/0550-3213(79)90031-2} {\bibfield  {journal}
  {\bibinfo  {journal} {Nucl. Phys. B}\ }\textbf {\bibinfo {volume} {156}},\
  \bibinfo {pages} {269} (\bibinfo {year} {1979})}\BibitemShut {NoStop}%
\bibitem [{\citenamefont {Christ}\ \emph {et~al.}(2010)\citenamefont {Christ},
  \citenamefont {Dawson}, \citenamefont {Izubuchi}, \citenamefont {Jung},
  \citenamefont {Liu}, \citenamefont {Mawhinney}, \citenamefont {Sachrajda},
  \citenamefont {Soni},\ and\ \citenamefont {Zhou}}]{Christ:2010dd}%
  \BibitemOpen
  \bibfield  {author} {\bibinfo {author} {\bibfnamefont {N.~H.}\ \bibnamefont
  {Christ}}, \bibinfo {author} {\bibfnamefont {C.}~\bibnamefont {Dawson}},
  \bibinfo {author} {\bibfnamefont {T.}~\bibnamefont {Izubuchi}}, \bibinfo
  {author} {\bibfnamefont {C.}~\bibnamefont {Jung}}, \bibinfo {author}
  {\bibfnamefont {Q.}~\bibnamefont {Liu}}, \bibinfo {author} {\bibfnamefont
  {R.~D.}\ \bibnamefont {Mawhinney}}, \bibinfo {author} {\bibfnamefont {C.~T.}\
  \bibnamefont {Sachrajda}}, \bibinfo {author} {\bibfnamefont {A.}~\bibnamefont
  {Soni}},\ and\ \bibinfo {author} {\bibfnamefont {R.}~\bibnamefont {Zhou}},\
  }\bibfield  {title} {\bibinfo {title} {{The $\eta$ and $\eta^\prime$ mesons
  from Lattice QCD}},\ }\href {https://doi.org/10.1103/PhysRevLett.105.241601}
  {\bibfield  {journal} {\bibinfo  {journal} {Phys. Rev. Lett.}\ }\textbf
  {\bibinfo {volume} {105}},\ \bibinfo {pages} {241601} (\bibinfo {year}
  {2010})},\ \Eprint {https://arxiv.org/abs/1002.2999} {arXiv:1002.2999
  [hep-lat]} \BibitemShut {NoStop}%
\bibitem [{\citenamefont {Fukaya}\ \emph {et~al.}(2015)\citenamefont {Fukaya},
  \citenamefont {Aoki}, \citenamefont {Cossu}, \citenamefont {Hashimoto},
  \citenamefont {Kaneko},\ and\ \citenamefont {Noaki}}]{Fukaya:2015ara}%
  \BibitemOpen
  \bibfield  {author} {\bibinfo {author} {\bibfnamefont {H.}~\bibnamefont
  {Fukaya}}, \bibinfo {author} {\bibfnamefont {S.}~\bibnamefont {Aoki}},
  \bibinfo {author} {\bibfnamefont {G.}~\bibnamefont {Cossu}}, \bibinfo
  {author} {\bibfnamefont {S.}~\bibnamefont {Hashimoto}}, \bibinfo {author}
  {\bibfnamefont {T.}~\bibnamefont {Kaneko}},\ and\ \bibinfo {author}
  {\bibfnamefont {J.}~\bibnamefont {Noaki}} (\bibinfo {collaboration}
  {JLQCD}),\ }\bibfield  {title} {\bibinfo {title} {{$\eta^\prime$ meson mass
  from topological charge density correlator in QCD}},\ }\href
  {https://doi.org/10.1103/PhysRevD.92.111501} {\bibfield  {journal} {\bibinfo
  {journal} {Phys. Rev. D}\ }\textbf {\bibinfo {volume} {92}},\ \bibinfo
  {pages} {111501} (\bibinfo {year} {2015})},\ \Eprint
  {https://arxiv.org/abs/1509.00944} {arXiv:1509.00944 [hep-lat]} \BibitemShut
  {NoStop}%
\bibitem [{\citenamefont {Ottnad}\ \emph {et~al.}(2012)\citenamefont {Ottnad},
  \citenamefont {Michael}, \citenamefont {Reker}, \citenamefont {Urbach},
  \citenamefont {Michael}, \citenamefont {Reker},\ and\ \citenamefont
  {Urbach}}]{Ottnad:2012fv}%
  \BibitemOpen
  \bibfield  {author} {\bibinfo {author} {\bibfnamefont {K.}~\bibnamefont
  {Ottnad}}, \bibinfo {author} {\bibfnamefont {C.}~\bibnamefont {Michael}},
  \bibinfo {author} {\bibfnamefont {S.}~\bibnamefont {Reker}}, \bibinfo
  {author} {\bibfnamefont {C.}~\bibnamefont {Urbach}}, \bibinfo {author}
  {\bibfnamefont {C.}~\bibnamefont {Michael}}, \bibinfo {author} {\bibfnamefont
  {S.}~\bibnamefont {Reker}},\ and\ \bibinfo {author} {\bibfnamefont
  {C.}~\bibnamefont {Urbach}} (\bibinfo {collaboration} {ETM}),\ }\bibfield
  {title} {\bibinfo {title} {{$\eta$ and $\eta^{\prime}$ mesons from
  $N_f=2+1+1$ twisted mass lattice QCD}},\ }\href
  {https://doi.org/10.1007/JHEP11(2012)048} {\bibfield  {journal} {\bibinfo
  {journal} {JHEP}\ }\textbf {\bibinfo {volume} {11}},\ \bibinfo {pages}
  {048}},\ \Eprint {https://arxiv.org/abs/1206.6719} {arXiv:1206.6719
  [hep-lat]} \BibitemShut {NoStop}%
\bibitem [{\citenamefont {Ottnad}\ \emph {et~al.}(2015)\citenamefont {Ottnad},
  \citenamefont {Urbach},\ and\ \citenamefont {Zimmermann}}]{Ottnad:2015hva}%
  \BibitemOpen
  \bibfield  {author} {\bibinfo {author} {\bibfnamefont {K.}~\bibnamefont
  {Ottnad}}, \bibinfo {author} {\bibfnamefont {C.}~\bibnamefont {Urbach}},\
  and\ \bibinfo {author} {\bibfnamefont {F.}~\bibnamefont {Zimmermann}}
  (\bibinfo {collaboration} {ETM}),\ }\bibfield  {title} {\bibinfo {title} {{A
  mixed action analysis of $\eta$ and $\eta^{\prime}$ mesons}},\ }\href
  {https://doi.org/10.1016/j.nuclphysb.2015.05.001} {\bibfield  {journal}
  {\bibinfo  {journal} {Nucl. Phys. B}\ }\textbf {\bibinfo {volume} {896}},\
  \bibinfo {pages} {470} (\bibinfo {year} {2015})},\ \Eprint
  {https://arxiv.org/abs/1501.02645} {arXiv:1501.02645 [hep-lat]} \BibitemShut
  {NoStop}%
\bibitem [{\citenamefont {Ottnad}\ and\ \citenamefont
  {Urbach}(2018)}]{Ottnad:2017bjt}%
  \BibitemOpen
  \bibfield  {author} {\bibinfo {author} {\bibfnamefont {K.}~\bibnamefont
  {Ottnad}}\ and\ \bibinfo {author} {\bibfnamefont {C.}~\bibnamefont {Urbach}}
  (\bibinfo {collaboration} {ETM}),\ }\bibfield  {title} {\bibinfo {title}
  {{Flavor-singlet meson decay constants from $N_f=2+1+1$ twisted mass lattice
  QCD}},\ }\href {https://doi.org/10.1103/PhysRevD.97.054508} {\bibfield
  {journal} {\bibinfo  {journal} {Phys. Rev. D}\ }\textbf {\bibinfo {volume}
  {97}},\ \bibinfo {pages} {054508} (\bibinfo {year} {2018})},\ \Eprint
  {https://arxiv.org/abs/1710.07986} {arXiv:1710.07986 [hep-lat]} \BibitemShut
  {NoStop}%
\bibitem [{\citenamefont {Dudek}\ \emph {et~al.}(2013)\citenamefont {Dudek},
  \citenamefont {Edwards}, \citenamefont {Guo},\ and\ \citenamefont
  {Thomas}}]{Dudek:2013yja}%
  \BibitemOpen
  \bibfield  {author} {\bibinfo {author} {\bibfnamefont {J.~J.}\ \bibnamefont
  {Dudek}}, \bibinfo {author} {\bibfnamefont {R.~G.}\ \bibnamefont {Edwards}},
  \bibinfo {author} {\bibfnamefont {P.}~\bibnamefont {Guo}},\ and\ \bibinfo
  {author} {\bibfnamefont {C.~E.}\ \bibnamefont {Thomas}} (\bibinfo
  {collaboration} {Hadron Spectrum}),\ }\bibfield  {title} {\bibinfo {title}
  {{Toward the excited isoscalar meson spectrum from lattice QCD}},\ }\href
  {https://doi.org/10.1103/PhysRevD.88.094505} {\bibfield  {journal} {\bibinfo
  {journal} {Phys. Rev. D}\ }\textbf {\bibinfo {volume} {88}},\ \bibinfo
  {pages} {094505} (\bibinfo {year} {2013})},\ \Eprint
  {https://arxiv.org/abs/1309.2608} {arXiv:1309.2608 [hep-lat]} \BibitemShut
  {NoStop}%
\bibitem [{\citenamefont {Bali}\ \emph {et~al.}(2015)\citenamefont {Bali},
  \citenamefont {Collins}, \citenamefont {D\"urr},\ and\ \citenamefont
  {Kanamori}}]{Bali:2014pva}%
  \BibitemOpen
  \bibfield  {author} {\bibinfo {author} {\bibfnamefont {G.~S.}\ \bibnamefont
  {Bali}}, \bibinfo {author} {\bibfnamefont {S.}~\bibnamefont {Collins}},
  \bibinfo {author} {\bibfnamefont {S.}~\bibnamefont {D\"urr}},\ and\ \bibinfo
  {author} {\bibfnamefont {I.}~\bibnamefont {Kanamori}},\ }\bibfield  {title}
  {\bibinfo {title} {{$D_s \rightarrow \eta, \eta'$ semileptonic decay form
  factors with disconnected quark loop contributions}},\ }\href
  {https://doi.org/10.1103/PhysRevD.91.014503} {\bibfield  {journal} {\bibinfo
  {journal} {Phys. Rev. D}\ }\textbf {\bibinfo {volume} {91}},\ \bibinfo
  {pages} {014503} (\bibinfo {year} {2015})},\ \Eprint
  {https://arxiv.org/abs/1406.5449} {arXiv:1406.5449 [hep-lat]} \BibitemShut
  {NoStop}%
\bibitem [{\citenamefont {Bali}\ \emph {et~al.}(2021)\citenamefont {Bali},
  \citenamefont {Braun}, \citenamefont {Collins}, \citenamefont {Sch\"afer},\
  and\ \citenamefont {Simeth}}]{Bali:2021qem}%
  \BibitemOpen
  \bibfield  {author} {\bibinfo {author} {\bibfnamefont {G.~S.}\ \bibnamefont
  {Bali}}, \bibinfo {author} {\bibfnamefont {V.}~\bibnamefont {Braun}},
  \bibinfo {author} {\bibfnamefont {S.}~\bibnamefont {Collins}}, \bibinfo
  {author} {\bibfnamefont {A.}~\bibnamefont {Sch\"afer}},\ and\ \bibinfo
  {author} {\bibfnamefont {J.}~\bibnamefont {Simeth}} (\bibinfo {collaboration}
  {RQCD}),\ }\bibfield  {title} {\bibinfo {title} {{Masses and decay constants
  of the \ensuremath{\eta} and \ensuremath{\eta}' mesons from lattice QCD}},\
  }\href {https://doi.org/10.1007/JHEP08(2021)137} {\bibfield  {journal}
  {\bibinfo  {journal} {JHEP}\ }\textbf {\bibinfo {volume} {08}},\ \bibinfo
  {pages} {137}},\ \Eprint {https://arxiv.org/abs/2106.05398} {arXiv:2106.05398
  [hep-lat]} \BibitemShut {NoStop}%
\bibitem [{\citenamefont {Gregory}\ \emph {et~al.}(2012)\citenamefont
  {Gregory}, \citenamefont {Irving}, \citenamefont {Richards},\ and\
  \citenamefont {McNeile}}]{Gregory:2011sg}%
  \BibitemOpen
  \bibfield  {author} {\bibinfo {author} {\bibfnamefont {E.~B.}\ \bibnamefont
  {Gregory}}, \bibinfo {author} {\bibfnamefont {A.~C.}\ \bibnamefont {Irving}},
  \bibinfo {author} {\bibfnamefont {C.~M.}\ \bibnamefont {Richards}},\ and\
  \bibinfo {author} {\bibfnamefont {C.}~\bibnamefont {McNeile}} (\bibinfo
  {collaboration} {UKQCD}),\ }\bibfield  {title} {\bibinfo {title} {{A study of
  the $\eta$ and $\eta^{\prime}$ mesons with improved staggered fermions}},\
  }\href {https://doi.org/10.1103/PhysRevD.86.014504} {\bibfield  {journal}
  {\bibinfo  {journal} {Phys. Rev. D}\ }\textbf {\bibinfo {volume} {86}},\
  \bibinfo {pages} {014504} (\bibinfo {year} {2012})},\ \Eprint
  {https://arxiv.org/abs/1112.4384} {arXiv:1112.4384 [hep-lat]} \BibitemShut
  {NoStop}%
\bibitem [{\citenamefont {Durr}(2012)}]{Durr:2012te}%
  \BibitemOpen
  \bibfield  {author} {\bibinfo {author} {\bibfnamefont {S.}~\bibnamefont
  {Durr}},\ }\bibfield  {title} {\bibinfo {title} {{Physics of $\eta'$ with
  rooted staggered quarks}},\ }\href
  {https://doi.org/10.1103/PhysRevD.85.114503} {\bibfield  {journal} {\bibinfo
  {journal} {Phys. Rev. D}\ }\textbf {\bibinfo {volume} {85}},\ \bibinfo
  {pages} {114503} (\bibinfo {year} {2012})},\ \Eprint
  {https://arxiv.org/abs/1203.2560} {arXiv:1203.2560 [hep-lat]} \BibitemShut
  {NoStop}%
\bibitem [{\citenamefont {Donald}\ \emph {et~al.}(2011)\citenamefont {Donald},
  \citenamefont {Davies}, \citenamefont {Follana},\ and\ \citenamefont
  {Kronfeld}}]{Donald:2011if}%
  \BibitemOpen
  \bibfield  {author} {\bibinfo {author} {\bibfnamefont {G.~C.}\ \bibnamefont
  {Donald}}, \bibinfo {author} {\bibfnamefont {C.~T.~H.}\ \bibnamefont
  {Davies}}, \bibinfo {author} {\bibfnamefont {E.}~\bibnamefont {Follana}},\
  and\ \bibinfo {author} {\bibfnamefont {A.~S.}\ \bibnamefont {Kronfeld}},\
  }\bibfield  {title} {\bibinfo {title} {{Staggered fermions, zero modes, and
  flavor-singlet mesons}},\ }\href {https://doi.org/10.1103/PhysRevD.84.054504}
  {\bibfield  {journal} {\bibinfo  {journal} {Phys. Rev. D}\ }\textbf {\bibinfo
  {volume} {84}},\ \bibinfo {pages} {054504} (\bibinfo {year} {2011})},\
  \Eprint {https://arxiv.org/abs/1106.2412} {arXiv:1106.2412 [hep-lat]}
  \BibitemShut {NoStop}%
\bibitem [{\citenamefont {Kogut}\ and\ \citenamefont
  {Susskind}(1975)}]{Kogut:1974ag}%
  \BibitemOpen
  \bibfield  {author} {\bibinfo {author} {\bibfnamefont {J.~B.}\ \bibnamefont
  {Kogut}}\ and\ \bibinfo {author} {\bibfnamefont {L.}~\bibnamefont
  {Susskind}},\ }\bibfield  {title} {\bibinfo {title} {{Hamiltonian Formulation
  of Wilson's Lattice Gauge Theories}},\ }\href
  {https://doi.org/10.1103/PhysRevD.11.395} {\bibfield  {journal} {\bibinfo
  {journal} {Phys. Rev. D}\ }\textbf {\bibinfo {volume} {11}},\ \bibinfo
  {pages} {395} (\bibinfo {year} {1975})}\BibitemShut {NoStop}%
\bibitem [{\citenamefont {Follana}\ \emph {et~al.}(2007)\citenamefont
  {Follana}, \citenamefont {Mason}, \citenamefont {Davies}, \citenamefont
  {Hornbostel}, \citenamefont {Lepage}, \citenamefont {Shigemitsu},
  \citenamefont {Trottier},\ and\ \citenamefont {Wong}}]{Follana:2006rc}%
  \BibitemOpen
  \bibfield  {author} {\bibinfo {author} {\bibfnamefont {E.}~\bibnamefont
  {Follana}}, \bibinfo {author} {\bibfnamefont {Q.}~\bibnamefont {Mason}},
  \bibinfo {author} {\bibfnamefont {C.}~\bibnamefont {Davies}}, \bibinfo
  {author} {\bibfnamefont {K.}~\bibnamefont {Hornbostel}}, \bibinfo {author}
  {\bibfnamefont {G.~P.}\ \bibnamefont {Lepage}}, \bibinfo {author}
  {\bibfnamefont {J.}~\bibnamefont {Shigemitsu}}, \bibinfo {author}
  {\bibfnamefont {H.}~\bibnamefont {Trottier}},\ and\ \bibinfo {author}
  {\bibfnamefont {K.}~\bibnamefont {Wong}} (\bibinfo {collaboration} {HPQCD,
  UKQCD}),\ }\bibfield  {title} {\bibinfo {title} {{Highly improved staggered
  quarks on the lattice, with applications to charm physics}},\ }\href
  {https://doi.org/10.1103/PhysRevD.75.054502} {\bibfield  {journal} {\bibinfo
  {journal} {Phys. Rev. D}\ }\textbf {\bibinfo {volume} {75}},\ \bibinfo
  {pages} {054502} (\bibinfo {year} {2007})},\ \Eprint
  {https://arxiv.org/abs/hep-lat/0610092} {arXiv:hep-lat/0610092} \BibitemShut
  {NoStop}%
\bibitem [{\citenamefont {Golterman}(2024)}]{Golterman:2024xos}%
  \BibitemOpen
  \bibfield  {author} {\bibinfo {author} {\bibfnamefont {M.}~\bibnamefont
  {Golterman}},\ }\bibfield  {title} {\bibinfo {title} {{Staggered fermions}},\
  }\href@noop {} {\  (\bibinfo {year} {2024})},\ \Eprint
  {https://arxiv.org/abs/2406.02906} {arXiv:2406.02906 [hep-lat]} \BibitemShut
  {NoStop}%
\bibitem [{\citenamefont {Sharpe}(2006)}]{Sharpe:2006re}%
  \BibitemOpen
  \bibfield  {author} {\bibinfo {author} {\bibfnamefont {S.~R.}\ \bibnamefont
  {Sharpe}},\ }\bibfield  {title} {\bibinfo {title} {{Rooted staggered
  fermions: Good, bad or ugly?}},\ }\href {https://doi.org/10.22323/1.032.0022}
  {\bibfield  {journal} {\bibinfo  {journal} {PoS}\ }\textbf {\bibinfo {volume}
  {LAT2006}},\ \bibinfo {pages} {022} (\bibinfo {year} {2006})},\ \Eprint
  {https://arxiv.org/abs/hep-lat/0610094} {arXiv:hep-lat/0610094} \BibitemShut
  {NoStop}%
\bibitem [{\citenamefont {Creutz}(2007{\natexlab{a}})}]{Creutz:2007rk}%
  \BibitemOpen
  \bibfield  {author} {\bibinfo {author} {\bibfnamefont {M.}~\bibnamefont
  {Creutz}},\ }\bibfield  {title} {\bibinfo {title} {{Why rooting fails}},\
  }\href {https://doi.org/10.22323/1.042.0007} {\bibfield  {journal} {\bibinfo
  {journal} {PoS}\ }\textbf {\bibinfo {volume} {LATTICE2007}},\ \bibinfo
  {pages} {007} (\bibinfo {year} {2007}{\natexlab{a}})},\ \Eprint
  {https://arxiv.org/abs/0708.1295} {arXiv:0708.1295 [hep-lat]} \BibitemShut
  {NoStop}%
\bibitem [{\citenamefont {Creutz}(2007{\natexlab{b}})}]{Creutz:2007yg}%
  \BibitemOpen
  \bibfield  {author} {\bibinfo {author} {\bibfnamefont {M.}~\bibnamefont
  {Creutz}},\ }\bibfield  {title} {\bibinfo {title} {{Chiral anomalies and
  rooted staggered fermions}},\ }\href
  {https://doi.org/10.1016/j.physletb.2007.03.065} {\bibfield  {journal}
  {\bibinfo  {journal} {Phys. Lett. B}\ }\textbf {\bibinfo {volume} {649}},\
  \bibinfo {pages} {230} (\bibinfo {year} {2007}{\natexlab{b}})},\ \Eprint
  {https://arxiv.org/abs/hep-lat/0701018} {arXiv:hep-lat/0701018} \BibitemShut
  {NoStop}%
\bibitem [{\citenamefont {Bernard}(2006)}]{Bernard:2006zw}%
  \BibitemOpen
  \bibfield  {author} {\bibinfo {author} {\bibfnamefont {C.}~\bibnamefont
  {Bernard}},\ }\bibfield  {title} {\bibinfo {title} {{Staggered chiral
  perturbation theory and the fourth-root trick}},\ }\href
  {https://doi.org/10.1103/PhysRevD.73.114503} {\bibfield  {journal} {\bibinfo
  {journal} {Phys. Rev. D}\ }\textbf {\bibinfo {volume} {73}},\ \bibinfo
  {pages} {114503} (\bibinfo {year} {2006})},\ \Eprint
  {https://arxiv.org/abs/hep-lat/0603011} {arXiv:hep-lat/0603011} \BibitemShut
  {NoStop}%
\bibitem [{\citenamefont {Kronfeld}(2007)}]{Kronfeld:2007ek}%
  \BibitemOpen
  \bibfield  {author} {\bibinfo {author} {\bibfnamefont {A.~S.}\ \bibnamefont
  {Kronfeld}},\ }\bibfield  {title} {\bibinfo {title} {{Lattice Gauge Theory
  with Staggered Fermions: How, Where, and Why (Not)}},\ }\href
  {https://doi.org/10.22323/1.042.0016} {\bibfield  {journal} {\bibinfo
  {journal} {PoS}\ }\textbf {\bibinfo {volume} {LATTICE2007}},\ \bibinfo
  {pages} {016} (\bibinfo {year} {2007})},\ \Eprint
  {https://arxiv.org/abs/0711.0699} {arXiv:0711.0699 [hep-lat]} \BibitemShut
  {NoStop}%
\bibitem [{\citenamefont {Golterman}(2008)}]{Golterman:2008gt}%
  \BibitemOpen
  \bibfield  {author} {\bibinfo {author} {\bibfnamefont {M.}~\bibnamefont
  {Golterman}},\ }\bibfield  {title} {\bibinfo {title} {{QCD with rooted
  staggered fermions}},\ }\href {https://doi.org/10.22323/1.077.0014}
  {\bibfield  {journal} {\bibinfo  {journal} {PoS}\ }\textbf {\bibinfo {volume}
  {CONFINEMENT8}},\ \bibinfo {pages} {014} (\bibinfo {year} {2008})},\ \Eprint
  {https://arxiv.org/abs/0812.3110} {arXiv:0812.3110 [hep-ph]} \BibitemShut
  {NoStop}%
\bibitem [{\citenamefont {Adams}(2008)}]{Adams:2008db}%
  \BibitemOpen
  \bibfield  {author} {\bibinfo {author} {\bibfnamefont {D.~H.}\ \bibnamefont
  {Adams}},\ }\bibfield  {title} {\bibinfo {title} {{The Rooting issue for a
  lattice fermion formulation similar to staggered fermions but without taste
  mixing}},\ }\href {https://doi.org/10.1103/PhysRevD.77.105024} {\bibfield
  {journal} {\bibinfo  {journal} {Phys. Rev. D}\ }\textbf {\bibinfo {volume}
  {77}},\ \bibinfo {pages} {105024} (\bibinfo {year} {2008})},\ \Eprint
  {https://arxiv.org/abs/0802.3029} {arXiv:0802.3029 [hep-lat]} \BibitemShut
  {NoStop}%
\bibitem [{\citenamefont {G\'erardin}\ \emph {et~al.}(2023)\citenamefont
  {G\'erardin}, \citenamefont {Verplanke}, \citenamefont {Wang}, \citenamefont
  {Fodor}, \citenamefont {Guenther}, \citenamefont {Lellouch}, \citenamefont
  {Szabo},\ and\ \citenamefont {Varnhorst}}]{Gerardin:2023naa}%
  \BibitemOpen
  \bibfield  {author} {\bibinfo {author} {\bibfnamefont {A.}~\bibnamefont
  {G\'erardin}}, \bibinfo {author} {\bibfnamefont {W.~E.~A.}\ \bibnamefont
  {Verplanke}}, \bibinfo {author} {\bibfnamefont {G.}~\bibnamefont {Wang}},
  \bibinfo {author} {\bibfnamefont {Z.}~\bibnamefont {Fodor}}, \bibinfo
  {author} {\bibfnamefont {J.~N.}\ \bibnamefont {Guenther}}, \bibinfo {author}
  {\bibfnamefont {L.}~\bibnamefont {Lellouch}}, \bibinfo {author}
  {\bibfnamefont {K.~K.}\ \bibnamefont {Szabo}},\ and\ \bibinfo {author}
  {\bibfnamefont {L.}~\bibnamefont {Varnhorst}},\ }\bibfield  {title} {\bibinfo
  {title} {{Lattice calculation of the $\pi^0$, $\eta$ and $\eta^{\prime}$
  transition form factors and the hadronic light-by-light contribution to the
  muon $g-2$}},\ }\href@noop {} {\  (\bibinfo {year} {2023})},\ \Eprint
  {https://arxiv.org/abs/2305.04570} {arXiv:2305.04570 [hep-lat]} \BibitemShut
  {NoStop}%
\bibitem [{\citenamefont {G\'erardin}(2021)}]{Gerardin:2020gpp}%
  \BibitemOpen
  \bibfield  {author} {\bibinfo {author} {\bibfnamefont {A.}~\bibnamefont
  {G\'erardin}},\ }\bibfield  {title} {\bibinfo {title} {{The anomalous
  magnetic moment of the muon: status of Lattice QCD calculations}},\ }\href
  {https://doi.org/10.1140/epja/s10050-021-00426-7} {\bibfield  {journal}
  {\bibinfo  {journal} {Eur. Phys. J. A}\ }\textbf {\bibinfo {volume} {57}},\
  \bibinfo {pages} {116} (\bibinfo {year} {2021})},\ \Eprint
  {https://arxiv.org/abs/2012.03931} {arXiv:2012.03931 [hep-lat]} \BibitemShut
  {NoStop}%
\bibitem [{\citenamefont {Aoyama}\ \emph {et~al.}(2020)\citenamefont {Aoyama}
  \emph {et~al.}}]{Aoyama:2020ynm}%
  \BibitemOpen
  \bibfield  {author} {\bibinfo {author} {\bibfnamefont {T.}~\bibnamefont
  {Aoyama}} \emph {et~al.},\ }\bibfield  {title} {\bibinfo {title} {{The
  anomalous magnetic moment of the muon in the Standard Model}},\ }\href
  {https://doi.org/10.1016/j.physrep.2020.07.006} {\bibfield  {journal}
  {\bibinfo  {journal} {Phys. Rept.}\ }\textbf {\bibinfo {volume} {887}},\
  \bibinfo {pages} {1} (\bibinfo {year} {2020})},\ \Eprint
  {https://arxiv.org/abs/2006.04822} {arXiv:2006.04822 [hep-ph]} \BibitemShut
  {NoStop}%
\bibitem [{\citenamefont {Navas}\ \emph {et~al.}(2024)\citenamefont {Navas}
  \emph {et~al.}}]{ParticleDataGroup:2024cfk}%
  \BibitemOpen
  \bibfield  {author} {\bibinfo {author} {\bibfnamefont {S.}~\bibnamefont
  {Navas}} \emph {et~al.} (\bibinfo {collaboration} {Particle Data Group}),\
  }\bibfield  {title} {\bibinfo {title} {{Review of particle physics}},\ }\href
  {https://doi.org/10.1103/PhysRevD.110.030001} {\bibfield  {journal} {\bibinfo
   {journal} {Phys. Rev. D}\ }\textbf {\bibinfo {volume} {110}},\ \bibinfo
  {pages} {030001} (\bibinfo {year} {2024})}\BibitemShut {NoStop}%
\bibitem [{\citenamefont {Michael}\ \emph {et~al.}(2013)\citenamefont
  {Michael}, \citenamefont {Ottnad},\ and\ \citenamefont
  {Urbach}}]{Michael:2013gka}%
  \BibitemOpen
  \bibfield  {author} {\bibinfo {author} {\bibfnamefont {C.}~\bibnamefont
  {Michael}}, \bibinfo {author} {\bibfnamefont {K.}~\bibnamefont {Ottnad}},\
  and\ \bibinfo {author} {\bibfnamefont {C.}~\bibnamefont {Urbach}} (\bibinfo
  {collaboration} {ETM}),\ }\bibfield  {title} {\bibinfo {title} {{$\eta$ and
  $\eta^\prime$ mixing from Lattice QCD}},\ }\href
  {https://doi.org/10.1103/PhysRevLett.111.181602} {\bibfield  {journal}
  {\bibinfo  {journal} {Phys. Rev. Lett.}\ }\textbf {\bibinfo {volume} {111}},\
  \bibinfo {pages} {181602} (\bibinfo {year} {2013})},\ \Eprint
  {https://arxiv.org/abs/1310.1207} {arXiv:1310.1207 [hep-lat]} \BibitemShut
  {NoStop}%
\bibitem [{\citenamefont {Aoki}\ \emph {et~al.}(2007)\citenamefont {Aoki},
  \citenamefont {Fukaya}, \citenamefont {Hashimoto},\ and\ \citenamefont
  {Onogi}}]{Aoki:2007ka}%
  \BibitemOpen
  \bibfield  {author} {\bibinfo {author} {\bibfnamefont {S.}~\bibnamefont
  {Aoki}}, \bibinfo {author} {\bibfnamefont {H.}~\bibnamefont {Fukaya}},
  \bibinfo {author} {\bibfnamefont {S.}~\bibnamefont {Hashimoto}},\ and\
  \bibinfo {author} {\bibfnamefont {T.}~\bibnamefont {Onogi}},\ }\bibfield
  {title} {\bibinfo {title} {{Finite volume QCD at fixed topological charge}},\
  }\href {https://doi.org/10.1103/PhysRevD.76.054508} {\bibfield  {journal}
  {\bibinfo  {journal} {Phys. Rev. D}\ }\textbf {\bibinfo {volume} {76}},\
  \bibinfo {pages} {054508} (\bibinfo {year} {2007})},\ \Eprint
  {https://arxiv.org/abs/0707.0396} {arXiv:0707.0396 [hep-lat]} \BibitemShut
  {NoStop}%
\bibitem [{\citenamefont {Borsanyi}\ \emph {et~al.}(2021)\citenamefont
  {Borsanyi} \emph {et~al.}}]{Borsanyi:2020mff}%
  \BibitemOpen
  \bibfield  {author} {\bibinfo {author} {\bibfnamefont {S.}~\bibnamefont
  {Borsanyi}} \emph {et~al.},\ }\bibfield  {title} {\bibinfo {title} {{Leading
  hadronic contribution to the muon magnetic moment from lattice QCD}},\ }\href
  {https://doi.org/10.1038/s41586-021-03418-1} {\bibfield  {journal} {\bibinfo
  {journal} {Nature}\ }\textbf {\bibinfo {volume} {593}},\ \bibinfo {pages}
  {51} (\bibinfo {year} {2021})},\ \Eprint {https://arxiv.org/abs/2002.12347}
  {arXiv:2002.12347 [hep-lat]} \BibitemShut {NoStop}%
\bibitem [{\citenamefont {Borsanyi}\ \emph {et~al.}(2018)\citenamefont
  {Borsanyi} \emph {et~al.}}]{Budapest-Marseille-Wuppertal:2017okr}%
  \BibitemOpen
  \bibfield  {author} {\bibinfo {author} {\bibfnamefont {S.}~\bibnamefont
  {Borsanyi}} \emph {et~al.} (\bibinfo {collaboration}
  {Budapest-Marseille-Wuppertal}),\ }\bibfield  {title} {\bibinfo {title}
  {{Hadronic vacuum polarization contribution to the anomalous magnetic moments
  of leptons from first principles}},\ }\href
  {https://doi.org/10.1103/PhysRevLett.121.022002} {\bibfield  {journal}
  {\bibinfo  {journal} {Phys. Rev. Lett.}\ }\textbf {\bibinfo {volume} {121}},\
  \bibinfo {pages} {022002} (\bibinfo {year} {2018})},\ \Eprint
  {https://arxiv.org/abs/1711.04980} {arXiv:1711.04980 [hep-lat]} \BibitemShut
  {NoStop}%
\bibitem [{\citenamefont {Altmeyer}\ \emph {et~al.}(1993)\citenamefont
  {Altmeyer}, \citenamefont {Born}, \citenamefont {Gockeler}, \citenamefont
  {Horsley}, \citenamefont {Laermann},\ and\ \citenamefont
  {Schierholz}}]{Altmeyer:1992dd}%
  \BibitemOpen
  \bibfield  {author} {\bibinfo {author} {\bibfnamefont {R.}~\bibnamefont
  {Altmeyer}}, \bibinfo {author} {\bibfnamefont {K.~D.}\ \bibnamefont {Born}},
  \bibinfo {author} {\bibfnamefont {M.}~\bibnamefont {Gockeler}}, \bibinfo
  {author} {\bibfnamefont {R.}~\bibnamefont {Horsley}}, \bibinfo {author}
  {\bibfnamefont {E.}~\bibnamefont {Laermann}},\ and\ \bibinfo {author}
  {\bibfnamefont {G.}~\bibnamefont {Schierholz}} (\bibinfo {collaboration}
  {MT(c)}),\ }\bibfield  {title} {\bibinfo {title} {{The Hadron spectrum in QCD
  with dynamical staggered fermions}},\ }\href
  {https://doi.org/10.1016/0550-3213(93)90328-M} {\bibfield  {journal}
  {\bibinfo  {journal} {Nucl. Phys. B}\ }\textbf {\bibinfo {volume} {389}},\
  \bibinfo {pages} {445} (\bibinfo {year} {1993})}\BibitemShut {NoStop}%
\bibitem [{\citenamefont {DeGrand}\ and\ \citenamefont
  {Schaefer}(2005)}]{DeGrand:2004wh}%
  \BibitemOpen
  \bibfield  {author} {\bibinfo {author} {\bibfnamefont {T.~A.}\ \bibnamefont
  {DeGrand}}\ and\ \bibinfo {author} {\bibfnamefont {S.}~\bibnamefont
  {Schaefer}},\ }\bibfield  {title} {\bibinfo {title} {{Improving meson
  two-point functions by low-mode averaging}},\ }\href
  {https://doi.org/10.1016/j.nuclphysbps.2004.11.352} {\bibfield  {journal}
  {\bibinfo  {journal} {Nucl. Phys. B Proc. Suppl.}\ }\textbf {\bibinfo
  {volume} {140}},\ \bibinfo {pages} {296} (\bibinfo {year} {2005})},\ \Eprint
  {https://arxiv.org/abs/hep-lat/0409056} {arXiv:hep-lat/0409056} \BibitemShut
  {NoStop}%
\bibitem [{\citenamefont {Giusti}\ \emph {et~al.}(2004)\citenamefont {Giusti},
  \citenamefont {Hernandez}, \citenamefont {Laine}, \citenamefont {Weisz},\
  and\ \citenamefont {Wittig}}]{Giusti:2004yp}%
  \BibitemOpen
  \bibfield  {author} {\bibinfo {author} {\bibfnamefont {L.}~\bibnamefont
  {Giusti}}, \bibinfo {author} {\bibfnamefont {P.}~\bibnamefont {Hernandez}},
  \bibinfo {author} {\bibfnamefont {M.}~\bibnamefont {Laine}}, \bibinfo
  {author} {\bibfnamefont {P.}~\bibnamefont {Weisz}},\ and\ \bibinfo {author}
  {\bibfnamefont {H.}~\bibnamefont {Wittig}},\ }\bibfield  {title} {\bibinfo
  {title} {{Low-energy couplings of QCD from current correlators near the
  chiral limit}},\ }\href {https://doi.org/10.1088/1126-6708/2004/04/013}
  {\bibfield  {journal} {\bibinfo  {journal} {JHEP}\ }\textbf {\bibinfo
  {volume} {04}},\ \bibinfo {pages} {013}},\ \Eprint
  {https://arxiv.org/abs/hep-lat/0402002} {arXiv:hep-lat/0402002} \BibitemShut
  {NoStop}%
\bibitem [{\citenamefont {Bali}\ \emph {et~al.}(2010)\citenamefont {Bali},
  \citenamefont {Collins},\ and\ \citenamefont {Schafer}}]{Bali:2009hu}%
  \BibitemOpen
  \bibfield  {author} {\bibinfo {author} {\bibfnamefont {G.~S.}\ \bibnamefont
  {Bali}}, \bibinfo {author} {\bibfnamefont {S.}~\bibnamefont {Collins}},\ and\
  \bibinfo {author} {\bibfnamefont {A.}~\bibnamefont {Schafer}},\ }\bibfield
  {title} {\bibinfo {title} {{Effective noise reduction techniques for
  disconnected loops in Lattice QCD}},\ }\href
  {https://doi.org/10.1016/j.cpc.2010.05.008} {\bibfield  {journal} {\bibinfo
  {journal} {Comput. Phys. Commun.}\ }\textbf {\bibinfo {volume} {181}},\
  \bibinfo {pages} {1570} (\bibinfo {year} {2010})},\ \Eprint
  {https://arxiv.org/abs/0910.3970} {arXiv:0910.3970 [hep-lat]} \BibitemShut
  {NoStop}%
\bibitem [{\citenamefont {Blum}\ \emph {et~al.}(2013)\citenamefont {Blum},
  \citenamefont {Izubuchi},\ and\ \citenamefont {Shintani}}]{Blum:2012uh}%
  \BibitemOpen
  \bibfield  {author} {\bibinfo {author} {\bibfnamefont {T.}~\bibnamefont
  {Blum}}, \bibinfo {author} {\bibfnamefont {T.}~\bibnamefont {Izubuchi}},\
  and\ \bibinfo {author} {\bibfnamefont {E.}~\bibnamefont {Shintani}},\
  }\bibfield  {title} {\bibinfo {title} {{New class of variance-reduction
  techniques using lattice symmetries}},\ }\href
  {https://doi.org/10.1103/PhysRevD.88.094503} {\bibfield  {journal} {\bibinfo
  {journal} {Phys. Rev. D}\ }\textbf {\bibinfo {volume} {88}},\ \bibinfo
  {pages} {094503} (\bibinfo {year} {2013})},\ \Eprint
  {https://arxiv.org/abs/1208.4349} {arXiv:1208.4349 [hep-lat]} \BibitemShut
  {NoStop}%
\bibitem [{\citenamefont {Shintani}\ \emph {et~al.}(2015)\citenamefont
  {Shintani}, \citenamefont {Arthur}, \citenamefont {Blum}, \citenamefont
  {Izubuchi}, \citenamefont {Jung},\ and\ \citenamefont
  {Lehner}}]{Shintani:2014vja}%
  \BibitemOpen
  \bibfield  {author} {\bibinfo {author} {\bibfnamefont {E.}~\bibnamefont
  {Shintani}}, \bibinfo {author} {\bibfnamefont {R.}~\bibnamefont {Arthur}},
  \bibinfo {author} {\bibfnamefont {T.}~\bibnamefont {Blum}}, \bibinfo {author}
  {\bibfnamefont {T.}~\bibnamefont {Izubuchi}}, \bibinfo {author}
  {\bibfnamefont {C.}~\bibnamefont {Jung}},\ and\ \bibinfo {author}
  {\bibfnamefont {C.}~\bibnamefont {Lehner}},\ }\bibfield  {title} {\bibinfo
  {title} {{Covariant approximation averaging}},\ }\href
  {https://doi.org/10.1103/PhysRevD.91.114511} {\bibfield  {journal} {\bibinfo
  {journal} {Phys. Rev. D}\ }\textbf {\bibinfo {volume} {91}},\ \bibinfo
  {pages} {114511} (\bibinfo {year} {2015})},\ \Eprint
  {https://arxiv.org/abs/1402.0244} {arXiv:1402.0244 [hep-lat]} \BibitemShut
  {NoStop}%
\bibitem [{\citenamefont {Venkataraman}\ and\ \citenamefont
  {Kilcup}(1997)}]{Venkataraman:1997xi}%
  \BibitemOpen
  \bibfield  {author} {\bibinfo {author} {\bibfnamefont {L.}~\bibnamefont
  {Venkataraman}}\ and\ \bibinfo {author} {\bibfnamefont {G.}~\bibnamefont
  {Kilcup}},\ }\bibfield  {title} {\bibinfo {title} {{The eta-prime meson with
  staggered fermions}},\ }\href@noop {} {\  (\bibinfo {year} {1997})},\ \Eprint
  {https://arxiv.org/abs/hep-lat/9711006} {arXiv:hep-lat/9711006} \BibitemShut
  {NoStop}%
\bibitem [{\citenamefont {Gregory}\ \emph {et~al.}(2008)\citenamefont
  {Gregory}, \citenamefont {Irving}, \citenamefont {Richards},\ and\
  \citenamefont {McNeile}}]{Gregory:2007ev}%
  \BibitemOpen
  \bibfield  {author} {\bibinfo {author} {\bibfnamefont {E.~B.}\ \bibnamefont
  {Gregory}}, \bibinfo {author} {\bibfnamefont {A.~C.}\ \bibnamefont {Irving}},
  \bibinfo {author} {\bibfnamefont {C.~M.}\ \bibnamefont {Richards}},\ and\
  \bibinfo {author} {\bibfnamefont {C.}~\bibnamefont {McNeile}},\ }\bibfield
  {title} {\bibinfo {title} {{Methods for Pseudoscalar Flavour-Singlet Mesons
  with Staggered Fermions}},\ }\href
  {https://doi.org/10.1103/PhysRevD.77.065019} {\bibfield  {journal} {\bibinfo
  {journal} {Phys. Rev. D}\ }\textbf {\bibinfo {volume} {77}},\ \bibinfo
  {pages} {065019} (\bibinfo {year} {2008})},\ \Eprint
  {https://arxiv.org/abs/0709.4224} {arXiv:0709.4224 [hep-lat]} \BibitemShut
  {NoStop}%
\bibitem [{\citenamefont {Giusti}\ \emph {et~al.}(2019)\citenamefont {Giusti},
  \citenamefont {Harris}, \citenamefont {Nada},\ and\ \citenamefont
  {Schaefer}}]{Giusti:2019kff}%
  \BibitemOpen
  \bibfield  {author} {\bibinfo {author} {\bibfnamefont {L.}~\bibnamefont
  {Giusti}}, \bibinfo {author} {\bibfnamefont {T.}~\bibnamefont {Harris}},
  \bibinfo {author} {\bibfnamefont {A.}~\bibnamefont {Nada}},\ and\ \bibinfo
  {author} {\bibfnamefont {S.}~\bibnamefont {Schaefer}},\ }\bibfield  {title}
  {\bibinfo {title} {{Frequency-splitting estimators of single-propagator
  traces}},\ }\href {https://doi.org/10.1140/epjc/s10052-019-7049-0} {\bibfield
   {journal} {\bibinfo  {journal} {Eur. Phys. J. C}\ }\textbf {\bibinfo
  {volume} {79}},\ \bibinfo {pages} {586} (\bibinfo {year} {2019})},\ \Eprint
  {https://arxiv.org/abs/1903.10447} {arXiv:1903.10447 [hep-lat]} \BibitemShut
  {NoStop}%
\bibitem [{\citenamefont {Lee}\ and\ \citenamefont
  {Sharpe}(1999)}]{Lee:1999zxa}%
  \BibitemOpen
  \bibfield  {author} {\bibinfo {author} {\bibfnamefont {W.-J.}\ \bibnamefont
  {Lee}}\ and\ \bibinfo {author} {\bibfnamefont {S.~R.}\ \bibnamefont
  {Sharpe}},\ }\bibfield  {title} {\bibinfo {title} {{Partial flavor symmetry
  restoration for chiral staggered fermions}},\ }\href
  {https://doi.org/10.1103/PhysRevD.60.114503} {\bibfield  {journal} {\bibinfo
  {journal} {Phys. Rev. D}\ }\textbf {\bibinfo {volume} {60}},\ \bibinfo
  {pages} {114503} (\bibinfo {year} {1999})},\ \Eprint
  {https://arxiv.org/abs/hep-lat/9905023} {arXiv:hep-lat/9905023} \BibitemShut
  {NoStop}%
\bibitem [{\citenamefont {Bernard}(2002)}]{Bernard:2001yj}%
  \BibitemOpen
  \bibfield  {author} {\bibinfo {author} {\bibfnamefont {C.}~\bibnamefont
  {Bernard}} (\bibinfo {collaboration} {MILC}),\ }\bibfield  {title} {\bibinfo
  {title} {{Chiral logs in the presence of staggered flavor symmetry
  breaking}},\ }\href {https://doi.org/10.1103/PhysRevD.65.054031} {\bibfield
  {journal} {\bibinfo  {journal} {Phys. Rev. D}\ }\textbf {\bibinfo {volume}
  {65}},\ \bibinfo {pages} {054031} (\bibinfo {year} {2002})},\ \Eprint
  {https://arxiv.org/abs/hep-lat/0111051} {arXiv:hep-lat/0111051} \BibitemShut
  {NoStop}%
\bibitem [{\citenamefont {Aubin}\ and\ \citenamefont
  {Bernard}(2003)}]{Aubin:2003mg}%
  \BibitemOpen
  \bibfield  {author} {\bibinfo {author} {\bibfnamefont {C.}~\bibnamefont
  {Aubin}}\ and\ \bibinfo {author} {\bibfnamefont {C.}~\bibnamefont
  {Bernard}},\ }\bibfield  {title} {\bibinfo {title} {{Pion and kaon masses in
  staggered chiral perturbation theory}},\ }\href
  {https://doi.org/10.1103/PhysRevD.68.034014} {\bibfield  {journal} {\bibinfo
  {journal} {Phys. Rev. D}\ }\textbf {\bibinfo {volume} {68}},\ \bibinfo
  {pages} {034014} (\bibinfo {year} {2003})},\ \Eprint
  {https://arxiv.org/abs/hep-lat/0304014} {arXiv:hep-lat/0304014} \BibitemShut
  {NoStop}%
\bibitem [{\citenamefont {Bernard}\ \emph {et~al.}(2001)\citenamefont
  {Bernard}, \citenamefont {Burch}, \citenamefont {Orginos}, \citenamefont
  {Toussaint}, \citenamefont {DeGrand}, \citenamefont {Detar}, \citenamefont
  {Datta}, \citenamefont {Gottlieb}, \citenamefont {Heller},\ and\
  \citenamefont {Sugar}}]{Bernard:2001av}%
  \BibitemOpen
  \bibfield  {author} {\bibinfo {author} {\bibfnamefont {C.~W.}\ \bibnamefont
  {Bernard}}, \bibinfo {author} {\bibfnamefont {T.}~\bibnamefont {Burch}},
  \bibinfo {author} {\bibfnamefont {K.}~\bibnamefont {Orginos}}, \bibinfo
  {author} {\bibfnamefont {D.}~\bibnamefont {Toussaint}}, \bibinfo {author}
  {\bibfnamefont {T.~A.}\ \bibnamefont {DeGrand}}, \bibinfo {author}
  {\bibfnamefont {C.~E.}\ \bibnamefont {Detar}}, \bibinfo {author}
  {\bibfnamefont {S.}~\bibnamefont {Datta}}, \bibinfo {author} {\bibfnamefont
  {S.~A.}\ \bibnamefont {Gottlieb}}, \bibinfo {author} {\bibfnamefont {U.~M.}\
  \bibnamefont {Heller}},\ and\ \bibinfo {author} {\bibfnamefont
  {R.}~\bibnamefont {Sugar}},\ }\bibfield  {title} {\bibinfo {title} {{The QCD
  spectrum with three quark flavors}},\ }\href
  {https://doi.org/10.1103/PhysRevD.64.054506} {\bibfield  {journal} {\bibinfo
  {journal} {Phys. Rev. D}\ }\textbf {\bibinfo {volume} {64}},\ \bibinfo
  {pages} {054506} (\bibinfo {year} {2001})},\ \Eprint
  {https://arxiv.org/abs/hep-lat/0104002} {arXiv:hep-lat/0104002} \BibitemShut
  {NoStop}%
\bibitem [{\citenamefont {Aubin}\ \emph {et~al.}(2004)\citenamefont {Aubin},
  \citenamefont {Bernard}, \citenamefont {DeTar}, \citenamefont {Osborn},
  \citenamefont {Gottlieb}, \citenamefont {Gregory}, \citenamefont {Toussaint},
  \citenamefont {Heller}, \citenamefont {Hetrick},\ and\ \citenamefont
  {Sugar}}]{Aubin:2004wf}%
  \BibitemOpen
  \bibfield  {author} {\bibinfo {author} {\bibfnamefont {C.}~\bibnamefont
  {Aubin}}, \bibinfo {author} {\bibfnamefont {C.}~\bibnamefont {Bernard}},
  \bibinfo {author} {\bibfnamefont {C.}~\bibnamefont {DeTar}}, \bibinfo
  {author} {\bibfnamefont {J.}~\bibnamefont {Osborn}}, \bibinfo {author}
  {\bibfnamefont {S.}~\bibnamefont {Gottlieb}}, \bibinfo {author}
  {\bibfnamefont {E.~B.}\ \bibnamefont {Gregory}}, \bibinfo {author}
  {\bibfnamefont {D.}~\bibnamefont {Toussaint}}, \bibinfo {author}
  {\bibfnamefont {U.~M.}\ \bibnamefont {Heller}}, \bibinfo {author}
  {\bibfnamefont {J.~E.}\ \bibnamefont {Hetrick}},\ and\ \bibinfo {author}
  {\bibfnamefont {R.}~\bibnamefont {Sugar}},\ }\bibfield  {title} {\bibinfo
  {title} {{Light hadrons with improved staggered quarks: Approaching the
  continuum limit}},\ }\href {https://doi.org/10.1103/PhysRevD.70.094505}
  {\bibfield  {journal} {\bibinfo  {journal} {Phys. Rev. D}\ }\textbf {\bibinfo
  {volume} {70}},\ \bibinfo {pages} {094505} (\bibinfo {year} {2004})},\
  \Eprint {https://arxiv.org/abs/hep-lat/0402030} {arXiv:hep-lat/0402030}
  \BibitemShut {NoStop}%
\bibitem [{\citenamefont {Aubin}\ \emph {et~al.}(2022)\citenamefont {Aubin},
  \citenamefont {Blum}, \citenamefont {Golterman},\ and\ \citenamefont
  {Peris}}]{Aubin:2022hgm}%
  \BibitemOpen
  \bibfield  {author} {\bibinfo {author} {\bibfnamefont {C.}~\bibnamefont
  {Aubin}}, \bibinfo {author} {\bibfnamefont {T.}~\bibnamefont {Blum}},
  \bibinfo {author} {\bibfnamefont {M.}~\bibnamefont {Golterman}},\ and\
  \bibinfo {author} {\bibfnamefont {S.}~\bibnamefont {Peris}},\ }\bibfield
  {title} {\bibinfo {title} {{Muon anomalous magnetic moment with staggered
  fermions: Is the lattice spacing small enough?}},\ }\href
  {https://doi.org/10.1103/PhysRevD.106.054503} {\bibfield  {journal} {\bibinfo
   {journal} {Phys. Rev. D}\ }\textbf {\bibinfo {volume} {106}},\ \bibinfo
  {pages} {054503} (\bibinfo {year} {2022})},\ \Eprint
  {https://arxiv.org/abs/2204.12256} {arXiv:2204.12256 [hep-lat]} \BibitemShut
  {NoStop}%
\bibitem [{\citenamefont {Bernard}\ \emph {et~al.}(2007)\citenamefont
  {Bernard}, \citenamefont {DeTar}, \citenamefont {Fu},\ and\ \citenamefont
  {Prelovsek}}]{Bernard:2007qf}%
  \BibitemOpen
  \bibfield  {author} {\bibinfo {author} {\bibfnamefont {C.}~\bibnamefont
  {Bernard}}, \bibinfo {author} {\bibfnamefont {C.~E.}\ \bibnamefont {DeTar}},
  \bibinfo {author} {\bibfnamefont {Z.}~\bibnamefont {Fu}},\ and\ \bibinfo
  {author} {\bibfnamefont {S.}~\bibnamefont {Prelovsek}},\ }\bibfield  {title}
  {\bibinfo {title} {{Scalar meson spectroscopy with lattice staggered
  fermions}},\ }\href {https://doi.org/10.1103/PhysRevD.76.094504} {\bibfield
  {journal} {\bibinfo  {journal} {Phys. Rev. D}\ }\textbf {\bibinfo {volume}
  {76}},\ \bibinfo {pages} {094504} (\bibinfo {year} {2007})},\ \Eprint
  {https://arxiv.org/abs/0707.2402} {arXiv:0707.2402 [hep-lat]} \BibitemShut
  {NoStop}%
\bibitem [{\citenamefont {Blossier}\ \emph {et~al.}(2009)\citenamefont
  {Blossier}, \citenamefont {Della~Morte}, \citenamefont {von Hippel},
  \citenamefont {Mendes},\ and\ \citenamefont {Sommer}}]{Blossier:2009kd}%
  \BibitemOpen
  \bibfield  {author} {\bibinfo {author} {\bibfnamefont {B.}~\bibnamefont
  {Blossier}}, \bibinfo {author} {\bibfnamefont {M.}~\bibnamefont
  {Della~Morte}}, \bibinfo {author} {\bibfnamefont {G.}~\bibnamefont {von
  Hippel}}, \bibinfo {author} {\bibfnamefont {T.}~\bibnamefont {Mendes}},\ and\
  \bibinfo {author} {\bibfnamefont {R.}~\bibnamefont {Sommer}},\ }\bibfield
  {title} {\bibinfo {title} {{On the generalized eigenvalue method for energies
  and matrix elements in lattice field theory}},\ }\href
  {https://doi.org/10.1088/1126-6708/2009/04/094} {\bibfield  {journal}
  {\bibinfo  {journal} {JHEP}\ }\textbf {\bibinfo {volume} {04}},\ \bibinfo
  {pages} {094}},\ \Eprint {https://arxiv.org/abs/0902.1265} {arXiv:0902.1265
  [hep-lat]} \BibitemShut {NoStop}%
\bibitem [{\citenamefont {Jay}\ and\ \citenamefont {Neil}(2021)}]{Jay:2020jkz}%
  \BibitemOpen
  \bibfield  {author} {\bibinfo {author} {\bibfnamefont {W.~I.}\ \bibnamefont
  {Jay}}\ and\ \bibinfo {author} {\bibfnamefont {E.~T.}\ \bibnamefont {Neil}},\
  }\bibfield  {title} {\bibinfo {title} {{Bayesian model averaging for analysis
  of lattice field theory results}},\ }\href
  {https://doi.org/10.1103/PhysRevD.103.114502} {\bibfield  {journal} {\bibinfo
   {journal} {Phys. Rev. D}\ }\textbf {\bibinfo {volume} {103}},\ \bibinfo
  {pages} {114502} (\bibinfo {year} {2021})},\ \Eprint
  {https://arxiv.org/abs/2008.01069} {arXiv:2008.01069 [stat.ME]} \BibitemShut
  {NoStop}%
\bibitem [{\citenamefont {G\'erardin}\ \emph {et~al.}(2019)\citenamefont
  {G\'erardin}, \citenamefont {Meyer},\ and\ \citenamefont
  {Nyffeler}}]{Gerardin:2019vio}%
  \BibitemOpen
  \bibfield  {author} {\bibinfo {author} {\bibfnamefont {A.}~\bibnamefont
  {G\'erardin}}, \bibinfo {author} {\bibfnamefont {H.~B.}\ \bibnamefont
  {Meyer}},\ and\ \bibinfo {author} {\bibfnamefont {A.}~\bibnamefont
  {Nyffeler}},\ }\bibfield  {title} {\bibinfo {title} {{Lattice calculation of
  the pion transition form factor with $N_f=2+1$ Wilson quarks}},\ }\href
  {https://doi.org/10.1103/PhysRevD.100.034520} {\bibfield  {journal} {\bibinfo
   {journal} {Phys. Rev. D}\ }\textbf {\bibinfo {volume} {100}},\ \bibinfo
  {pages} {034520} (\bibinfo {year} {2019})},\ \Eprint
  {https://arxiv.org/abs/1903.09471} {arXiv:1903.09471 [hep-lat]} \BibitemShut
  {NoStop}%
\bibitem [{\citenamefont {Alexandrou}\ \emph
  {et~al.}(2023{\natexlab{a}})\citenamefont {Alexandrou} \emph
  {et~al.}}]{ExtendedTwistedMass:2023jan}%
  \BibitemOpen
  \bibfield  {author} {\bibinfo {author} {\bibfnamefont {C.}~\bibnamefont
  {Alexandrou}} \emph {et~al.} (\bibinfo {collaboration} {Extended Twisted
  Mass}),\ }\bibfield  {title} {\bibinfo {title}
  {{\ensuremath{\eta}\textrightarrow{}\ensuremath{\gamma}*\ensuremath{\gamma}*
  transition form factor and the hadronic light-by-light \ensuremath{\eta}-pole
  contribution to the muon g-2 from lattice QCD}},\ }\href
  {https://doi.org/10.1103/PhysRevD.108.054509} {\bibfield  {journal} {\bibinfo
   {journal} {Phys. Rev. D}\ }\textbf {\bibinfo {volume} {108}},\ \bibinfo
  {pages} {054509} (\bibinfo {year} {2023}{\natexlab{a}})}\BibitemShut
  {NoStop}%
\bibitem [{\citenamefont {Alexandrou}\ \emph
  {et~al.}(2023{\natexlab{b}})\citenamefont {Alexandrou} \emph
  {et~al.}}]{Alexandrou:2023lia}%
  \BibitemOpen
  \bibfield  {author} {\bibinfo {author} {\bibfnamefont {C.}~\bibnamefont
  {Alexandrou}} \emph {et~al.},\ }\bibfield  {title} {\bibinfo {title} {{Pion
  Transition Form Factor from Twisted-Mass Lattice QCD and the Hadronic
  Light-by-Light $\pi^0$-pole Contribution to the Muon $g-2$}},\ }\href@noop {}
  {\  (\bibinfo {year} {2023}{\natexlab{b}})},\ \Eprint
  {https://arxiv.org/abs/2308.12458} {arXiv:2308.12458 [hep-lat]} \BibitemShut
  {NoStop}%
\end{thebibliography}%

\end{document}